\documentclass[10pt,twocolumn,twoside]{IEEEtran}

\IEEEoverridecommandlockouts     

\newcommand{\blue}[1]{{\color{blue} #1}}   

\usepackage{color,amsmath,amstext,amsfonts,amssymb, mathrsfs,mathtools}
\usepackage{graphicx,algorithm}
\usepackage[algo2e,linesnumbered]{algorithm2e}
\usepackage{tikz}
\usepackage{setspace}
\usepackage{array} 
\usepackage{booktabs}  
\usepackage{bbm}
\usepackage{wasysym}
\usepackage{textcomp}
\usepackage{hyperref} 
\usepackage{lipsum}
\usepackage{subfig}
\usepackage[sort,compress]{cite}

\newtheorem{lemma}{Lemma}
\newtheorem{assumption}{Assumption}

\newtheorem{theorem}{Theorem}
\newtheorem{definition}{Definition}
\newtheorem{corollary}{Corollary}

\newtheorem{remark}{Remark}

\SetKwInput{KwInput}{Input}

\newcounter{l1}
\newcounter{l2}
\newcounter{l3}
\newcommand{\bdotlist}{\begin{list}{$\bullet$}{}}
\newcommand{\bboxlist}{\begin{list}{$\Box$}{}}
\newcommand{\bbboxlist}{\begin{list}{\raisebox{.005in}{{\tiny $\blacksquare$ \ \ }}}{}}
\newcommand{\bdashlist}{\begin{list}{$-$}{} }
\newcommand{\blist}{\begin{list}{}{} }
\newcommand{\barablist}{\begin{list}{\arabic{l1}}{\usecounter{l1}}}
\newcommand{\balphlist}{\begin{list}{(\alph{l2})}{\usecounter{l2}}}
\newcommand{\bAlphlist}{\begin{list}{\Alph{l2}.}{\usecounter{l2}}}
\newcommand{\bdiamlist}{\begin{list}{$\diamond$}{}}
\newcommand{\bromalist}{\begin{list}{(\roman{l3})}{\usecounter{l3}}}

\newcommand{\beq}{\begin{equation}}
\newcommand{\eeq}{\end{equation}}

\title{Online Algorithms for Network Robustness under Connectivity Constraints}

\author{Deepan Muthirayan and Pramod P. Khargonekar
\thanks{This work is supported in part by the National Science Foundation under Grant ECCS-1839429.
D. Muthirayan and P. P. Khargonekar are with the Department of Electrical Engineering and Computer Sciences, University of California Irvine, Irvine, CA (emails: deepan.m@uci.edu, pramod.khargonekar@uci.edu).
}
}

\begin{document}

\maketitle
\thispagestyle{empty}
\pagestyle{empty}

\begin{abstract}
In this paper, we  present algorithms for designing networks that are robust to node failures with minimal or limited number of links. We present algorithms for both the static network setting and the dynamic network setting; setting where new nodes can arrive in the future. For the static setting, we present  algorithms for constructing the optimal network in terms of the number of links used for a given node size and the number of nodes that can fail. We then consider the dynamic setting where it is disruptive to remove any of the older links. For this setting, we present  online algorithms for two cases: (i) when the number of nodes that can fail remains constant and (ii) when only the proportion of the nodes that can fail remains constant. We show that the proposed algorithm for the first case saves nearly $3/4$th of the total possible links at any point of time. We then present algorithms for various levels of the fraction of the nodes that can fail and characterize their link usage. We show that when $1/2$ the number of nodes can fail at any point of time, the proposed algorithm saves nearly $1/2$ of the total possible links at any point of time. We show that when the number of nodes that can fail is limited to the fraction $1/(2m)$ ($m \in \mathbb{N}$), the proposed algorithm saves nearly as much as $(1-1/2m)$ of the total possible links at any point of time. We also show that when the number of nodes that can fail at any point of time is $1/2$ of the number of nodes plus $n$, $n \in \mathbb{N}$, the number of links saved by the proposed algorithm reduces only linearly in $n$. We conjecture that the saving ratio achieved by the algorithms we present is optimal for the dynamic setting.
\end{abstract}


\section{Introduction.}\label{intro}

Connectivity is a central objective in network applications. A road network uses roads to connect geographical locations along which people travel from one location to the other; a telecommunication network uses links and nodes to connect users. Thus, it is important that in a network, there should be at least one path connecting every location to every other location. Further, it is important that this connectivity property be robust to network disruptions. Real networks will typically have multiple objectives. For example, in a road network, the time of travel, the cost of the network, etc. are also very important. It is therefore likely that in practice, network design  will require some  optimization objective with connectivity as one of the primary constraints. Thus, the network solution in real networks would be a subset of the solutions that satisfy (robust) connectivity constraint. In this work, we study the problem of network optimization for this basic constraint: {\it connectivity}. Our goal is to design algorithms for this problem in the dynamic network setting.

A dynamic network is a network that changes with time. Examples of such networks can be readily found in nature, society, and  modern technological systems: molecular interactions, social networks, transportation networks, mobile wireless networks, power grids and robot collectives. 
These networks exhibit different types of dynamics, for example, changing nature of the nodes and links or changing number of nodes. Recent years have witnessed growing interest in theoretical frameworks for dynamic networks. To a great extent this is motivated by the changing landscape of the Internet, the rise of Internet of Things (IoT), and their accompanying technical challenges. For example, the number of mobile-only users has already exceeded the number of desktop-only users and more than 75\% of all digital consumers are now using both desktop and mobile platforms to access the Internet. The Internet of Things and smart cities are fast becoming a reality \cite{mone2015new}. All of these systems and their networks are inherently dynamic. 


Dynamic networks add a new dimension to network design and analysis – time. It is widely recognized that dynamic networks exhibit different structural and algorithmic properties \cite{aggarwal2014evolutionary}. In this work we consider the design of dynamic networks constantly changing by the addition of new nodes. The central question we address is how to update the network minimally such that the network continues to be robust. We study this problem under one central premise motivated by many physical networks: the network cannot be modified by disrupting the older portion of the network. To the best of our knowledge, this is the first paper to  study this problem from a theory perspective. Throughout this paper, we use network to refer to the underlying graph structure. 

Static networks (or graphs) and their theoretical properties have been extensively studied in the literature \cite{gibbons1985algorithmic, babai1995automorphism, godsil2001algebraic}. These works typically study the relation of properties such as {\it node connectivity}, $\kappa$ (the smallest number of nodes whose removal results in a disconnected network), the {\it link connectivity}, $\lambda$ (the smallest number of links whose removal results in a disconnected network), the minimum degree, $d_{\text{min}}$, and the number of node distinct paths and link distinct paths. In \cite{gibbons1985algorithmic}, it is shown that $\kappa \leq \lambda \leq d_{\text{min}}$ for a general network. 

The optimality condition for connectivity for a regular network is defined as $\kappa = \lambda = d$. In \cite{babai1995automorphism, godsil2001algebraic}, it is shown that $\lambda = d$ for node-similar regular graphs and $\kappa = \lambda = d$ for symmetric regular graphs. Dekker et al. \cite{dekker2004network} showed that the $q-$dimensional hypertorus satisfies $\kappa = \lambda = d = 2q$. Thus, the hypertorus can be used to design optimal networks. The challenge is that the application of this procedure for the generation of the optimal network for a given node size and $\kappa$ is not feasible for all node sizes and $\kappa$. 

Alternatively, group theoretic concepts can also be used to generate optimal networks \cite{dekker2004network, babai1995automorphism, biggs1993algebraic} for a general $\kappa$. Here the optimal graphs are generated by constructing the minimal Cayley graph for a set. This procedure can be non-trivial and complex because it requires the identification of a set of a suitable size and a minimal generator set for that set whose size matches $\kappa$. The noteworthy point here is the relation between optimal networks and groups and their minimal generators. 
Two-level network structures based on an optimal base-network and a hub-network can also be used to construct optimal networks which are not necessarily node-similar and thus provide more general generation procedure for optimal networks. This design is guaranteed to be optimal provided that there exists an automorphism between the nodes in the base-network that are connected to different nodes in the hub-network (see \cite{dekker2004network}). 

The dynamic setting that we consider is non-trivial because (i) the optimal networks for a lesser node size need not be a sub-network of the optimal network for a larger node size and (ii) $\kappa$ can be dynamic. The procedures given above for generating the optimal network for a given node size and $\kappa$ are (i) either not feasible or complex and (ii) not in a suitable form to extend to the dynamic setting. 


\subsection{Our Contributions}

In this work we {\it develop algorithms for designing networks that are optimal or limited in the number of links used while they are robust to node failures}. First, we present the static network setting. We develop {\it algorithms for constructing the optimal network in terms of the number of links used for a given node size and robustness, $N_f$ ($= \kappa - 1$), the number of nodes that can fail without disconnecting the network}. 
Then, we present the dynamic network setting. We develop algorithms for two cases: (i) when the number of nodes that can fail remains constant and (ii) when only the proportion of the nodes that can fail remains constant (changing $N_f$). 
We present algorithms that are robust to node failures at any point of time and characterize their link bounds for each of the cases. For the second case, we present algorithms for different levels of the fraction of the nodes that can fail.

We show that the proposed algorithm for the first case saves nearly $3/4$th of the total possible links at any point of time. When the fraction of the nodes that can fail at any point of time is $1/2$, we show that the proposed algorithm saves nearly $1/2$ of the total possible links at any point of time. In particular, when the fraction of the nodes that can fail is $1/(2m)$ ($m \in \mathbb{N}$), we show that the proposed algorithm saves nearly as much as $(1-1/2m)$ of the total possible links at any point of time. We also show that when the number of nodes that can fail at any point of time is $1/2$ of the number of nodes plus $n$, $n \in \mathbb{N}$, the number of links saved by the algorithm we propose reduces only linearly in $n$. {\it Our key contribution is presenting online algorithms for various levels of robustness for the dynamic setting}. 

\subsection{Related Work}

Many other graph theoretic properties have also been studied in the literature. Bollobas et al. \cite{bollobas2001random} showed that in the limit as the number of nodes goes to infinity a random graph (or network) is optimal. The {\it diameter}, defined as the longest of all the shortest paths is also an important graph-theoretic property. Bollobas et al. \cite{bollobas2001random} present general characterization of the diameter for regular and node-similar networks. Dekker et al. \cite{dekker2004network} also characterize the relation between {\it average distance} between nodes and the diameter for node-similar graphs. {\it Link load}, which quantifies a notion of traffic through a link, is another important graph-theoretic property. In \cite{dekker2004network}, characterization for the link load for regular and node-similar networks is presented. 

Many papers have also proposed models for generating networks that mimic properties of real-world networks. Watts and Strogatz \cite{watts1998collective} proposed a simple model to generate what are called `small world' networks that are characterized by high clustering and small average distance. Barabasi et al. \cite{barabasi1999emergence,  albert2002statistical,barabasi2003linked} introduced the {\it scale-free networks}. 
The distinctive feature of these networks is that they exhibit the power law degree distribution observed in many real world networks like the world wide web etc. 
But these networks are vulnerable to attacks. Dekker et al. \cite{dekker2004network} showed that the two-level network structure can be used to combine an optimal network and a scale-free network to generate networks that retain the benefits of scale-free networks and that are robust simultaneously. The Klemm and Eguiluz (KE) model proposed in \cite{klemm2002highly} can be used to generate networks that satisfy the power law degree distribution and the properties of small world networks. 

Several authors propose various metrics for measuring the robustness of networks. In \cite{crucitti2003efficiency, holme2002attack, latora2004science, criado2006new}, authors study robustness from the point of view of decrease of efficiency due to attacks. Criado et al. \cite{criado2005effective} propose an alternate metric called {\it vulnerability} for measuring the robustness of networks. Dekker et al. \cite{dekker2005symmetry} propose a metric called {\it symmetric ratio}. 
Manzano et al. \cite{manzano2013endurance} propose a measure called {\it endurance} as a flexible metric for evaluating network robustness under various failure scenarios. Piraveenan et al. \cite{piraveenan2013quantifying} propose a metric to measure robustness to targeted sustained attacks by measuring the changing size of the largest component as the network disintegrates. Schieber et al. \cite{schieber2016information} give an information theoretic perspective on network robustness. Bigdeli et al. \cite{bigdeli2009comparison} present extensive simulations to analyze the networks from the point of view connectivity, {\it network criticality}, average node degree and node betweeness. Liu et al. \cite{liu2017comparative} present a comparative study of various robustness metrics.

Many authors also study the robustness of networks in various applications and provide application specific metrics and procedures. Bilal et al.  \cite{bilal2013characterization} propose robustness metric for Data-Center Networks (DCN). Manzano et al. \cite{manzano2013connectivity} present robustness analysis of standard DCNs such as 1) ThreeTier, 2) FatTree and 3) DCell. Tizghadam et al. \cite{tizghadam2009autonomic} propose robust network control strategies based on network criticality for communication networks. Tizghadam et al. \cite{tizghadam2008robust} propose traffic routing in transportation networks based on the network criticality metric. Rueda et al. \cite{rueda2017robustness} empirically compare the robustness of 15 telecommunication networks. Chen et al. \cite{chen2017robustness} investigate the robustness of power-grid and communication networks to cascading failures. Wang et al. \cite{wang2017multi} empirically study the robustness of metro networks. In \cite{lordan2016robustness, zhou2019efficiency}, the authors empirically study the robustness of airline networks. Pallis et al. \cite{pallis2009structure} study the properties of real-world vehicular communication networks. Angskun \cite{angskun2007binomial} propose a scalable and fault-tolerant network topology for high performance computing. Yazicioglu et al. \cite{yaziciouglu2015formation} present a decentralized scheme for converting any multi-agent interaction graph by adding new links locally to improve the robustness of the interaction graph. Pederson et al. \cite{pedersen2006applying} propose extensions of the 4-layer grid structure for large scale multi-processor network systems for several criterions. Bellingeri et al. \cite{bellingeri2018robustness} empirically study the robustness properties of weighted networks. 

Generation methods for heterogenous networks have also been presented in the literature. Ramanathan et al. \cite{ramanathan2000topology}
propose the Local Information No Topology (LINT) and
Local Information Link-state Topology (LILT) algorithms for wireless communication networks. Borbash et al. \cite{borbash2002distributed} discuss a distributed relative neighborhood graph algorithm (Dist-RNG) that minimizes congestion while maintaining connectivity. Ning et al. \cite{li2004topology} propose the Directed Relative Neighborhood Graph (DRNG) and Directed Local Minimum Spanning Tree (DLMST) algorithms to enable wireless communication networks to reduce congestion. While the heterogeneous network generation procedures discussed above can lack robustness, \cite{venuturumilli2010obtaining} propose a distributed algorithm for generating robust heterogeneous networks. Rose et al. \cite{qiu2017rose} present a robust algorithm for wireless sensor networks. 



{\it Notation}: We denote a network by $\mathbf{G} = (\mathbf{N}, \mathbf{L})$, where $\mathbf{N}$ denotes the set of nodes and $\mathbf{L}$ denotes the adjacency matrix, where $\mathbf{L}^{i,j} = l^{i,j}, \ l^{i,j} = 1$ or $0$, $l^{i,i} = 0$. The size of $\mathbf{N}$ and $\mathbf{L}$ is denoted by $N = \vert \mathbf{N} \vert$ and $L = \sum_{i}\sum_{j > i} l^{i,j}$ respectively. 

\section{Problem Formulation}
\label{problemform.}

We denote the dynamic network as a sequence of networks $\{ \mathbf{G}_k, \ k = 0,1,....\}$ indexed by time. The network corresponding to time instant $k$, $\mathbf{G}_k$, is given by $\mathbf{G}_k = \left(\mathbf{N}_k, \mathbf{L}_k\right)$, where $\mathbf{N}_k$ denotes the set of nodes and $\mathbf{L}_k$ denotes the adjacency matrix corresponding to $\mathbf{G}_k$. We denote the corresponding number of nodes and links by $N_k$ and $L_k$. 

The number of nodes of the dynamic network evolves according to $N_{k+1} = N_k + m_k, \ m_k \in \mathbb{N}, \ m_k > 0$, where $m_k$ is arbitrary. We make the following assumptions.
\begin{assumption}
{\it Any pair of nodes can be connected by a link. All feasible links are of the same length.}
\label{ass:as1}
\end{assumption}
\begin{assumption}
{\it Once the links are formed they cannot be removed.}
\label{ass:as2}
\end{assumption}

Assumption \ref{ass:as1} is a standard assumption. Assumption \ref{ass:as2} is not a limiting assumption but rather an assumption that makes the dynamic setting challenging. By $\mathbf{G}_i \subset \mathbf{G}_j$ we imply that $\mathbf{G}_i$ is a sub-network of $\mathbf{G}_j$. By $\mathcal{G}^{N_f}$ we denote the set of all networks derived from $\mathbf{G}$ by removing any of the $N_f$ nodes of $\mathbf{G}$. We define the following constraints.

\begin{definition}
{\it ($N_f$ Robustness) $\mathbf{RG}_{N_f}$: $\bf{H}$ is connected $\forall \ \mathbf{H} \ \in \ \mathcal{G}^{N_f}$}
\label{def:robustness}
\end{definition}

\begin{definition}
{\it (Link Constraint) $\mathbf{LG}_N$: $L \leq \frac{N^2}{4} + N - 2$}
\label{def:linkconstraint}
\end{definition}

The action for the network manager at the instant $k$ is to decide which of the feasible links to add. Denote the decision to connect two nodes $i$ and $j$ by $u^{i,j}_k$. Then the decision options for the network manager are as follows:
\begin{equation}
u^{i,j}_k = \left\{ \begin{array}{cc} 1 & \ \text{form link between} \ i \ \text{and} \ j \\  0 & \text{no action between node} \ i \ \text{and} \ j \end{array}\right. . \nonumber
\end{equation}

We denote the overall decision by the matrix $U_k$, whose $(i,j)$th element is given by $u^{i,j}_k$. We state the dynamic network robustness problem as follows: {\it for $N_{k+1} = N_k + m_k, \ m_k \in \mathbb{N}, \ m_k > 0$, and under Assumption \ref{ass:as1}, make the sequence of decisions $\{U_k, k = 1,2,....\}$ such that
\begin{enumerate}
\item $\mathbf{G}_i \subseteq \mathbf{G}_j$ $\forall i,j \ \text{and} \ i < j$
\item Robustness $\mathbf{RG}_{N_{f,k}}$ holds $\forall \ k$
\item the number of links is minimal at any point of time
\end{enumerate}}

The problem poses the following question: {\it how to continuously update a growing network without disrupting the older portion such that the robustness constraint is satisfied at all times and the number links used is minimal}? The robustness condition requires that the network is robust to failure of $N_{f,k}$ nodes at the instant $k$, and thus can vary with time. 



\section{Static Network Setting}

We first discuss the static network setting. The static network problem can be stated as follows: {\it what is the minimal $L$ under Assumption \ref{ass:as1} such that the robustness constraint $\mathbf{RG}_{N_f}$ holds}? 
We note that without Assumption \ref{ass:as2} the dynamic network problem reduces to this problem. We define the optimal network as the network constructed with the minimal $L$ and that satisfies robustness. First we make a simple observation on the minimal connectivity that is required for every node in a network so that the network is robust. 

\begin{lemma}
{\it For any $N_f$, each node of a network of size $N$ should have at least $N_f+1$ links for the network to remain connected even after any of the $N_f$ nodes fails.}
\label{lem:minimal-edges}
\end{lemma}
{\em Proof}:
We prove by contradiction. Suppose there is a node $a$ which has only $N_f$ links. Then the number of nodes connected to node $a$ is $N_f$. It is clear that on removal of all the nodes connected to node $a$ the resulting network is not connected because then the node $a$ will become isolated. Hence, there exist a set of $N_f$ nodes which on removal makes the network disconnected. This proves the claim that each node should have atleast $N_f+1$ links. $\blacksquare$

This lemma will be used to prove the optimality of the algorithms we provide herein. In the next theorem we present the algorithm for constructing the optimal network for a general $N$ (even) and robustness $N_f (< N-1)$. The algorithm we present starts by connecting the nodes in a cycle, which we denote by $\tilde{\mathbf{G}}_2$ going forward (an example of cycle for $N = 8$ is shown in Fig. \ref{fig:net-8-7-f>1}), and then proceeds by connecting the neighbors that are within certain number of hops from each other in $\tilde{\mathbf{G}}_2$. 

\begin{theorem}
{\it For a network of $N$ nodes such that $N$ is even and any $N_f$ s.t. $N_f+1 = \tilde{N}_f < N$, the following algorithm: (i) construct a cycle $\tilde{\mathbf{G}}_2$ with the $N$ nodes (ii) When $\tilde{N}_f$ is even connect every node to the nodes that are within $\tilde{N}_f/2$ links in $\tilde{\mathbf{G}}_2$ and (iii) When $\tilde{N}_f$ is odd connect every node to the nodes that are within $(\tilde{N}_f-1)/2$ links in $\tilde{\mathbf{G}}_2$ and connect every pair of nodes that are $N/2$ links apart in $\tilde{\mathbf{G}}_2$ (diametrically opposite nodes); constructs the optimal network for robustness $N_f$.}
\label{thm:even-node-connection}
\end{theorem}
Please see Appendix \ref{sec:pf-Th1} for the proof. In Fig. \ref{fig:net-8-7-f>1}, we illustrate the network constructed according to Theorem \ref{thm:even-node-connection} for $N = 8$, and $N_f = 3, 4$. 

\subsection{Alternate Optimal Constructions}

In the theorems to follow we give alternate constructions for the optimal network when the robustness is specified by the factor $f (\geq 1)$ as $N_f = N/(2f)$, where $N/(2f) \in \mathbb{N}$. 
We will use the same factor $f$ to specify the fraction of the nodes that can fail in the dynamic setting. We present the construction for $f = 1$ and $f > 1$ as different results for better illustration of the ideas. The algorithms we present later in the dynamic setting 
draws its ideas from the construction presented in the theorems to follow. We first consider the $f = 1$ case. First, we characterize the minimum required connectivity for a network to be robust to failure of half of the nodes. 
\begin{lemma}
{\it For any network $\bf{G}$ of $N (N = 2m, \ m \in \mathbb{N}, \ m \geq 3)$ nodes and $L < L^1_{opt}(N) = \frac{N^2}{4} + \frac{N}{2}$ links $\exists$ a set of $\frac{N}{2}$ nodes which on removal results in a network that is not connected. Also $\exists$ a $\bf{G}$ s.t. $L = L^1_{opt}(N)$ and $\mathbf{RG}_{\frac{N}{2}}$ holds.}
\label{lem:static-rob-f-1}
\end{lemma}

Please see Appendix \ref{sec:pf-Lm2} for the proof. The proof establishes its conclusions by giving a construction that (i) {\it has two fully connected halves}, and (ii) {\it where each node in each half is connected to two other nodes in the other half such that no set of $n < N/2$ nodes in one half are connected to the same set of $n < N/2$ nodes in the other half}. 
In the next theorem we formally establish that these general rules generate the optimal network for robustness specified by $f=1$. 
\blue{\begin{figure*}
\subfloat{\includegraphics[width = 2.25in]{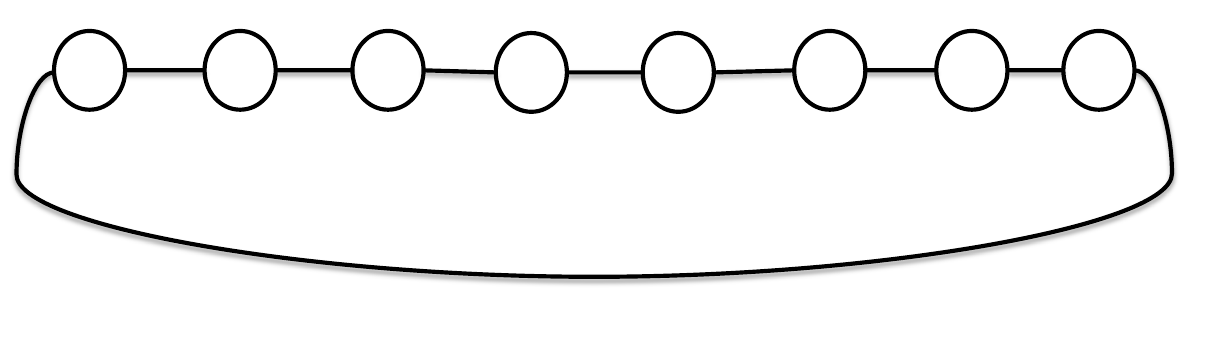}}
\subfloat{\includegraphics[width = 2.25in]{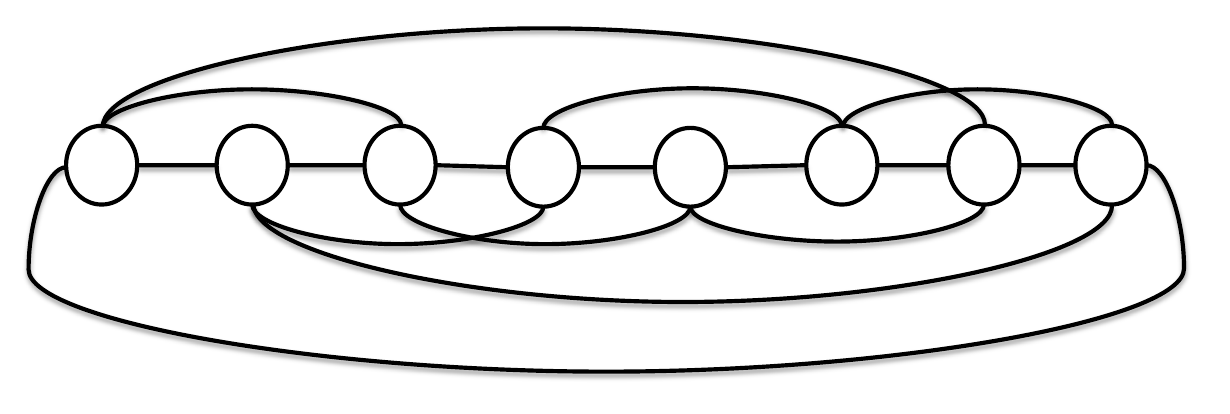}}
\subfloat{\includegraphics[width = 2.25in]{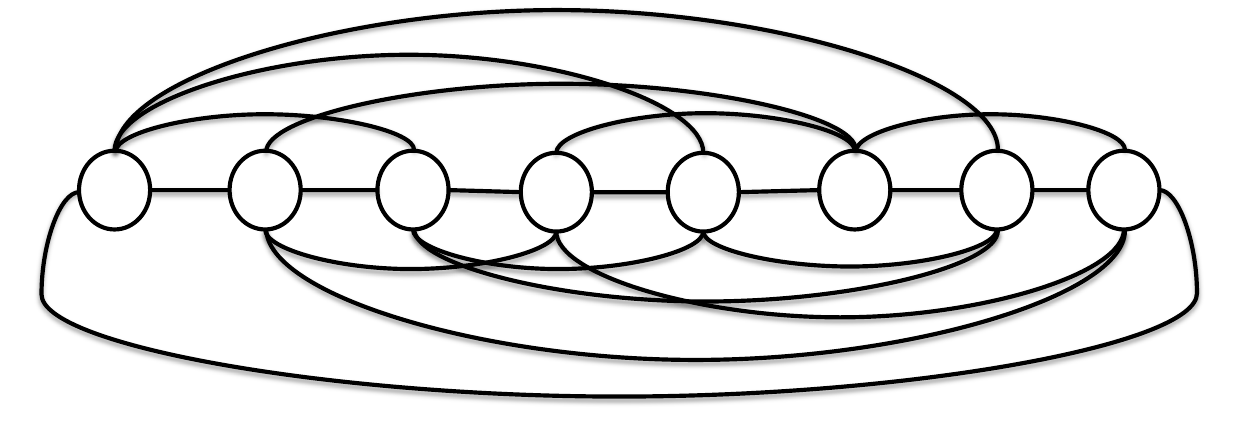}} \\
\subfloat{\includegraphics[width = 2.25in]{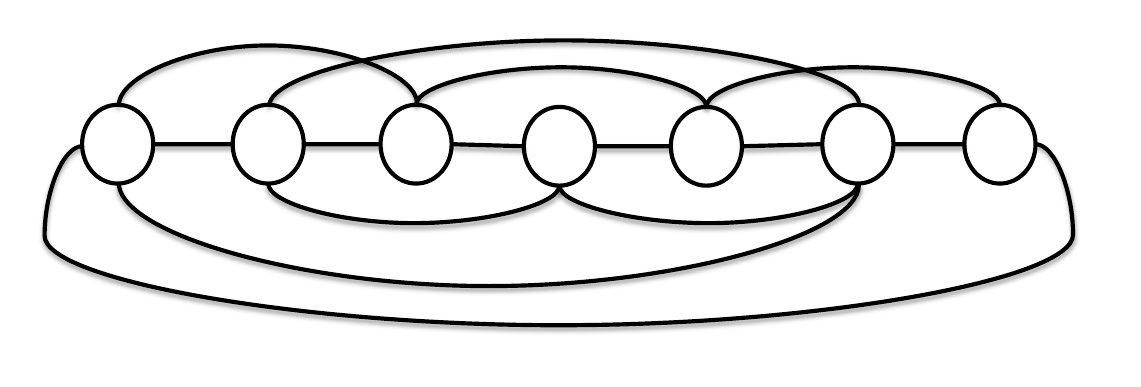}}
\subfloat{\includegraphics[width = 2.25in]{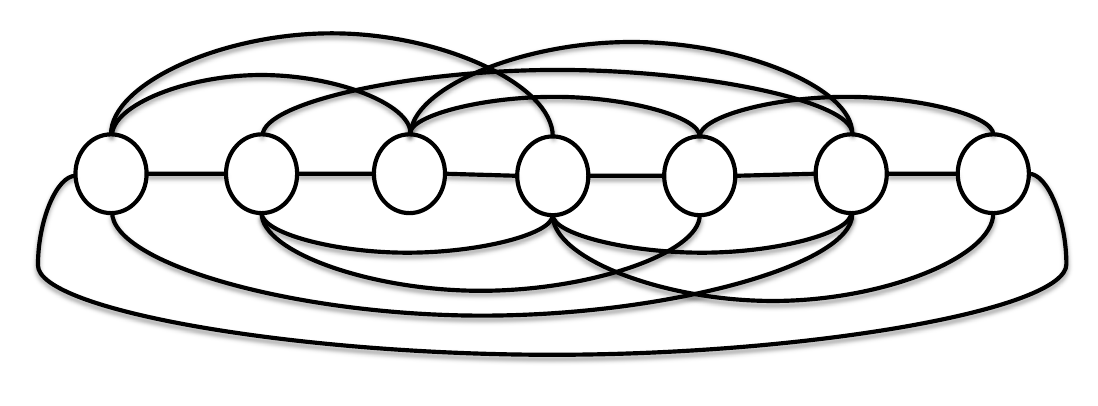}}
\caption{Top left: $\tilde{\mathbf{G}}_2$ for $N = 8$. Top middle: network for $N = 8$ and $N_f = 3$. Top right: network with $N = 8$ and $N_f = 4$. Bottom left: network with $N = 7$ and $N_f = 3$. Bottom right: network with $N = 7$ and $N_f = 4$.}
\label{fig:net-8-7-f>1} 
\end{figure*}}

\begin{figure}
\subfloat{\includegraphics[width = 1.7in]{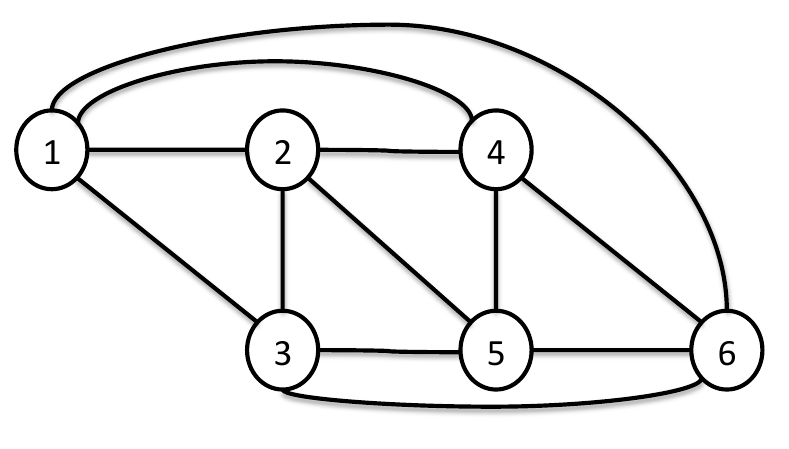}}
\subfloat{\includegraphics[width = 1.7in]{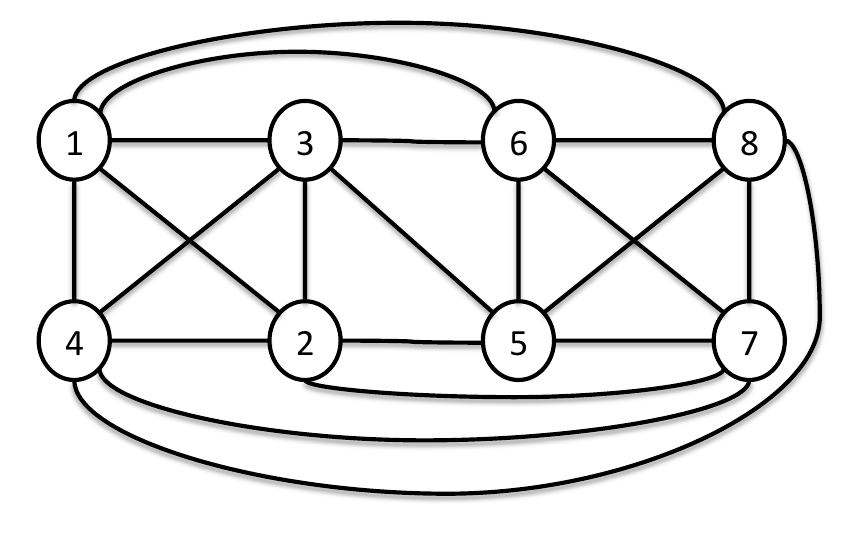}}
\caption{Optimal network for $f = 1$. Left: $N = 6$. Right: $N = 8$.}
\label{fig:net-6-8-f-1} 
\end{figure}

\begin{theorem}
{\it For $N (N = 2m, \ m \in \mathbb{N}, m \geq 3)$, the following algorithm (A1) : (i) \it construct two fully connected halves, (ii) connect each node in each half to two other nodes in the other half such that no set of $n < N/2$ nodes in one half are connected to the same set of $n < N/2$ nodes in the other half, constructs the optimal network for $f = 1$.}
\label{thm:alg-optimal-f-1}
\end{theorem}

Please see Appendix \ref{sec:pf-Th2} for the proof. Figures \ref{fig:net-6-8-f-1} shows the optimal network constructed according to Theorem \ref{thm:alg-optimal-f-1} for $N = 6$ and $N = 8$ nodes. In these two examples the nodes are numbered such that the corresponding adjacency matrix equals the construction $\mathbf{L}^{1,opt}_{N}$ given in the proof of Lemma \ref{lem:static-rob-f-1}. 

In the next theorem we present the algorithm that constructs the optimal network for robustness $f > 1$. First, we present an algorithm to construct a network that satisfies $N_f-$robustness for any $N_f < N-1$ when $N$ is odd. Similar to the algorithm prescribed in Theorem \ref{thm:even-node-connection}, this algorithm starts with the construction of a cycle $\tilde{\mathbf{G}}_2$ and then proceeds by connecting the nodes to its neighbors but with a minor difference. 

\begin{lemma}
{\it For a network of $N$ nodes such that $N$ is odd and any $N_f$ s.t. $N_f+1 = \tilde{N}_f < N$, the following algorithm: (i) construct a cycle $\tilde{\mathbf{G}}_2$ with the $N$ nodes (ii) When $\tilde{N}_f$ is even connect every node to the nodes that are within $\tilde{N}_f/2$ links in $\tilde{\mathbf{G}}_2$ and (iii)  When $\tilde{N}_f$ is odd connect every node to the nodes that are within $(\tilde{N}_f-1)/2$ links in $\tilde{\mathbf{G}}_2$ and connect every node to one of the node that is $(N-1)/2$ links away in $\tilde{\mathbf{G}}_2$ such that only one node is connected to $\tilde{N}_f+1$ nodes; constructs the network that satisfies $N_f$-robustness.}
\label{lem:odd-node-connection}
\end{lemma}
The proof starts by giving a construction that satisfies the step (iii) of the lemma. The argument to prove the robustness of this network is similar to the argument presented in Theorem \ref{thm:even-node-connection}. Please see Appendix \ref{sec:pf-Lm3} for the detailed proof. In Fig. \ref{fig:net-8-7-f>1}, we illustrate the network constructed according to Lemma \ref{lem:odd-node-connection} for $N = 7$, and $N_f = 3, 4$.   

Next, we present the algorithm that constructs the optimal network for a general $f > 1$. The construction we present is similar to the previous theorem and proceeds by splitting the nodes into two distinct halves. It then uses the construction presented in Theorem \ref{thm:even-node-connection} and Lemma \ref{lem:odd-node-connection} to connect the nodes within the two halves.
\begin{theorem}
{\it For any $N (N = 2m, \ m \in \mathbb{N}, m \geq 3), f > 1, N_f = N/(2f)$ the following algorithm (Af): (i) {\it split the nodes into two sets (halves) of $N/2$ nodes}, (ii) {\it connect each node in one half to a node in the other half such that every node in one half is connected to a different node in the other half}, (iii) {\it if $N/(2)$ is odd then connect the nodes in each half as in Lemma \ref{lem:odd-node-connection} by setting $\tilde{N}_f = N/(2f)$ such that the nodes connected to $\tilde{N}_f+1$ nodes in each half are connected to each other}, (iv) {\it else connect the nodes in each half as in Theorem \ref{thm:even-node-connection} by setting $\tilde{N}_f = N/(2f)$}, (v) {\it remove the link connecting the nodes connected to $\tilde{N}_f+1$ nodes within their respective halves}; constructs the optimal network for robustness $N_f$.}
\label{thm:alg-optimal-f}
\end{theorem}

\begin{figure}
\subfloat{\includegraphics[width = 1.5in]{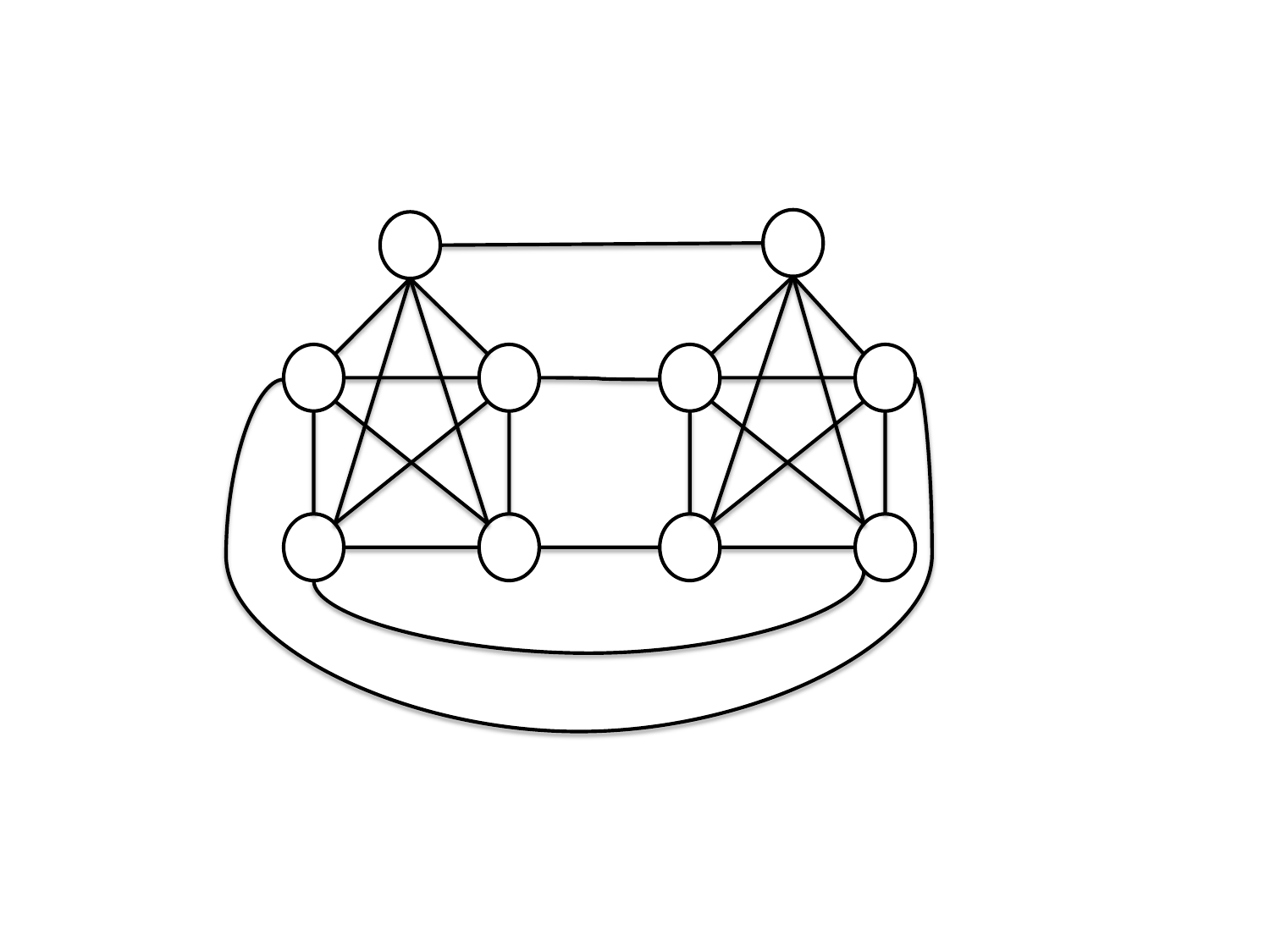}}
\subfloat{\includegraphics[width = 1.5in]{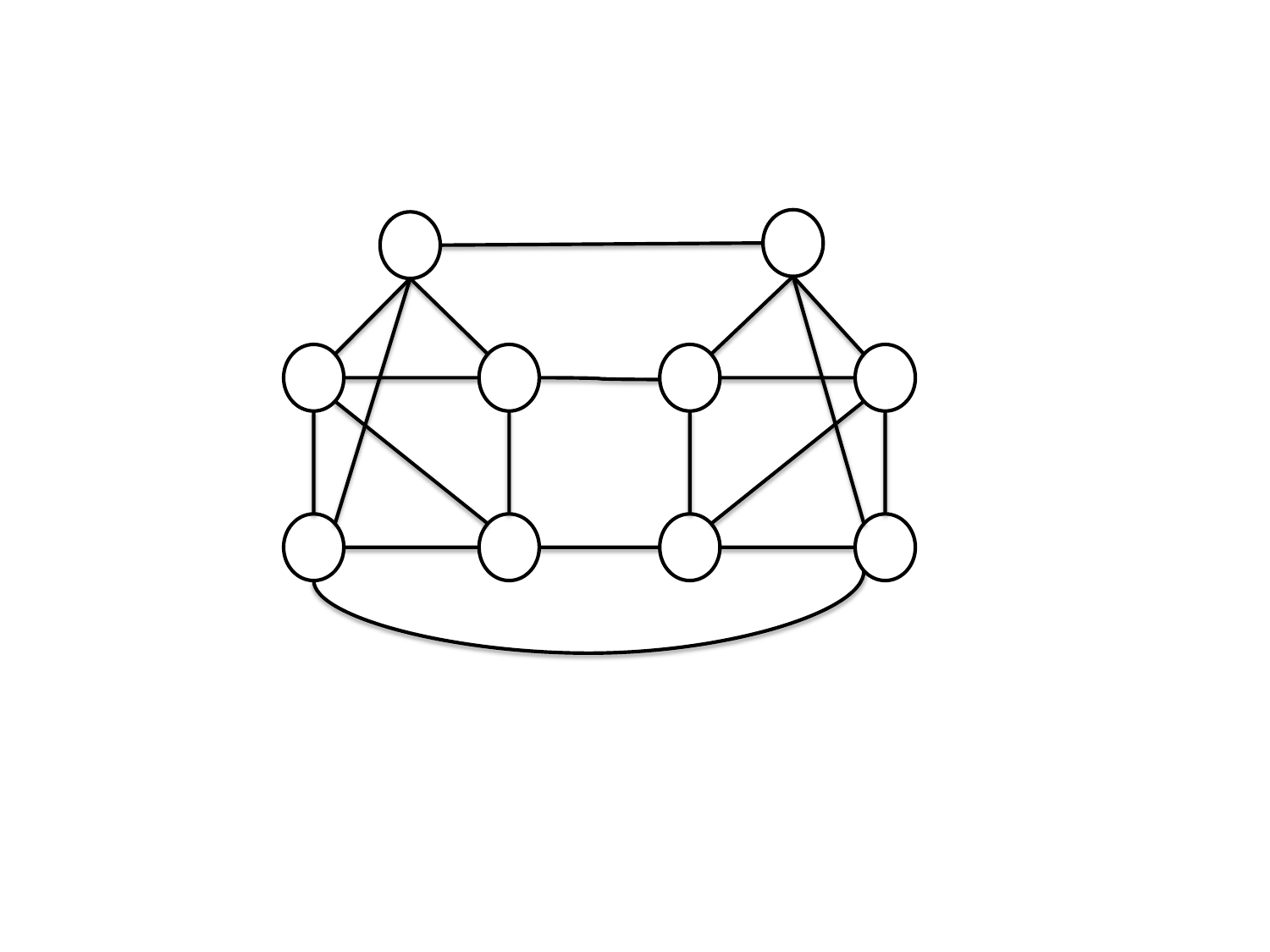}}\\
\subfloat{\includegraphics[width = 1.5in]{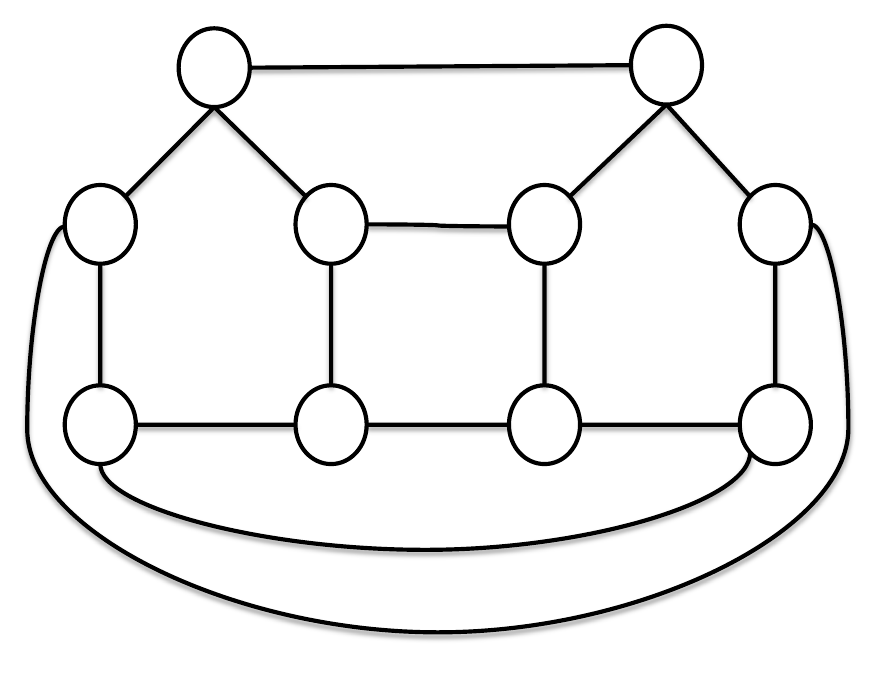}}
\subfloat{\includegraphics[width = 1.5in]{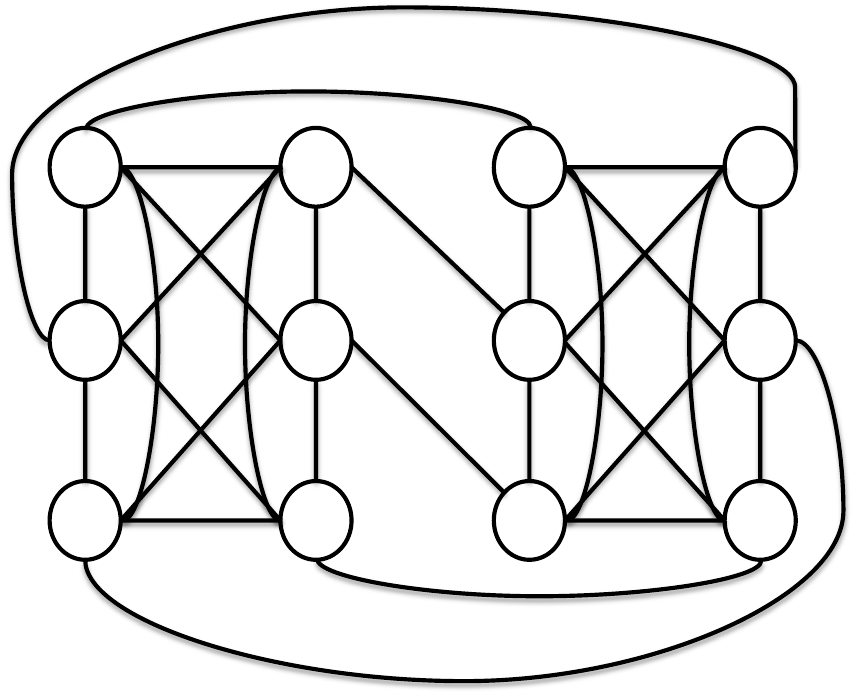}}\\
\subfloat{\includegraphics[width = 1.5in]{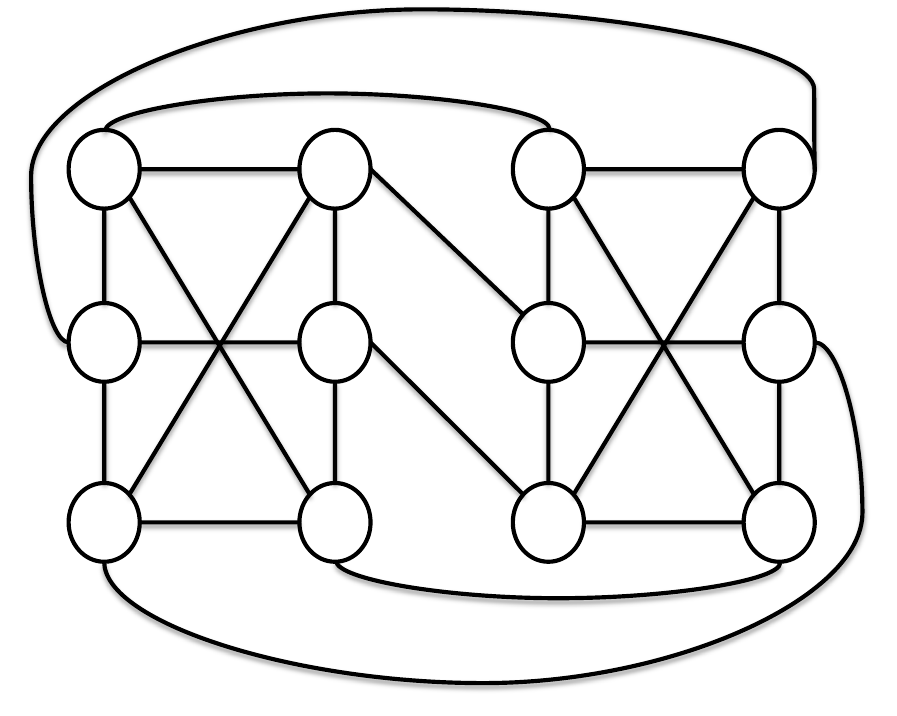}}
\subfloat{\includegraphics[width = 1.5in]{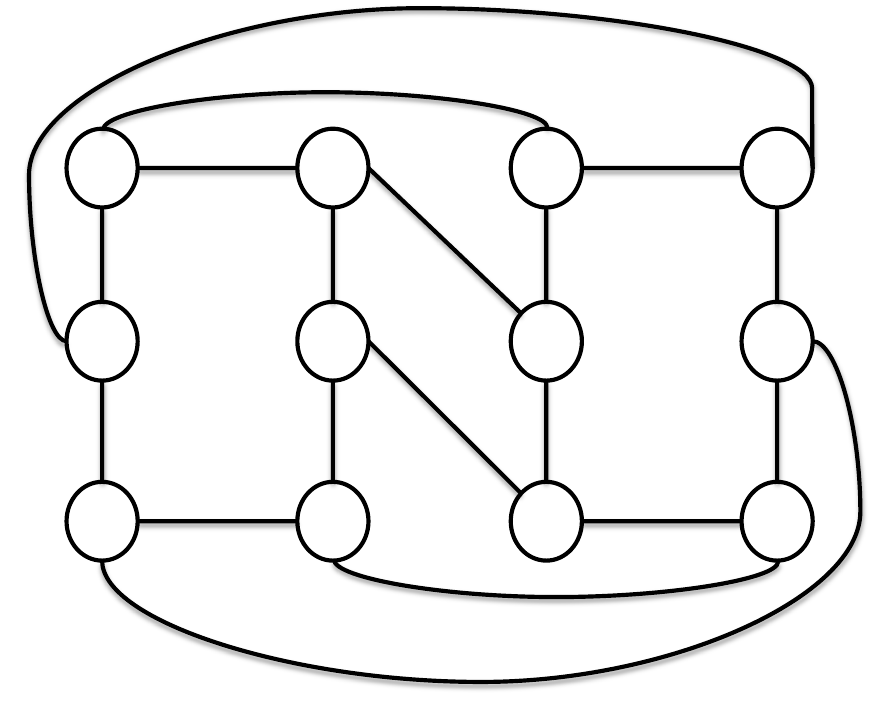}}
\caption{Top left: optimal 10 node network for $N_f = 4$. Top right: optimal 10 node network for $N_f = 3$. Middle left: optimal 10 node network for $N_f = 2$. Middle right: optimal 12 node network for $N_f = 4$. Bottom left: optimal 12 node network for $N_f = 3$. Bottom right: optimal 12 node network for $N_f = 2$.}
\label{fig:opt-net-10-12-f>1} 
\end{figure}

From Theorem \ref{thm:even-node-connection} and Lemma \ref{lem:odd-node-connection} it follows that each of the two halves of the construction in Theorem \ref{thm:alg-optimal-f} satisfy $(N_f - 1)$-robustness. The proof shows that even after removing any of the $N_f$ nodes there is always a pair of nodes with one from each of the two halves that is connected. It then follows that when less than $N_f$ nodes are removed from one half and the remaining from the other half the network will always remain connected. The construction is also such that when $N_f$ nodes are removed from one half each node in this half is connected to a node in the other half. Thus the network will always remain connected after failure of any of the $N_f$ nodes. Please see Appendix \ref{sec:pf-Th3} for the detailed proof.  

In the next theorem we extend the idea in Theorem \ref{thm:alg-optimal-f} to present an alternate construction for the optimal network for robustness $N/(2m)$, $m \in \mathbb{N}$, when $N$ is $2m$ divisible. The construction we present for this specific case is useful in specific dynamic settings we present later.
\begin{theorem}
{\it For $m \in \mathbb{N}, m > 1$, suppose $N$ is $2m$ divisible, then the following algorithm: (i) {\it split the nodes into $2m$ sets of $N/(2m)$ nodes}, (ii) {\it fully connect each set}, (iii) {\it form $2m$ cycles with a unique node from each set in each cycle}; constructs the optimal network for robustness $N/(2m)$.}
\label{thm:alg-optimal-m}
\end{theorem}

Please see Appendix \ref{sec:pf-Th4} for the proof. Figure \ref{fig:opt-net-10-12-f>1} illustrates the optimal network constructed according to Af for $N = 10, 12$ and $N_f = 2,3,4$ respectively. We note that there are distinctly two sets of $N/2$ nodes in each network. We also observe that each node in each network is connected to exactly $N_f+1$ links. In the next corollary we present the number of links required to construct the optimal network for a general $f \geq 1$.  
\begin{corollary}
{\it For any $f \geq 1$ s.t. $N_f = N/(2f) \in \mathbb{N}$, the number of links in the optimal network is $L^f_{opt} = N^2/4 +N/2 -Nm/2, \ m \in \mathbb{N}$, where $m = N/2-N_f$.}
\label{cor:opt-links-f}
\end{corollary}
The proof trivially follows from the observation that the link count is equal to the number of ones in the corresponding adjacency matrix divided by two. 

\begin{remark}
The algorithms presented in Theorem \ref{thm:even-node-connection}, Lemma \ref{lem:odd-node-connection}, Theorem \ref{thm:alg-optimal-f-1}, and Theorem \ref{thm:alg-optimal-f} have complexity characterized by $\mathcal{O}(N^2)$ and the algorithm in Theorem \ref{thm:alg-optimal-m} has complexity characterized by $\mathcal{O}(N^2/m^2+N)$.
\end{remark}

\section{Dynamic Network Setting}

In this section we present the algorithms for different scenarios of node  failures for the dynamic setting. We first present the case where the number of nodes that can fail is constant and then present the case where the fraction of nodes that can fail is constant. We present online algorithms with link bounds for each of the cases. Throughout, the robustness requirement, $N_f$, is taken to be equal to the number of nodes that can fail at any point of time. Hence, the robustness requirement is changing with time. The objective of the algorithms is to make connectivity decisions $U_k$ without disrupting the older part of the network such that the network remains robust and the number of links is minimal. 

\subsection{When the number of nodes that can fail is constant}

The difficulty here is that the procedures developed for the static case cannot be applied here. The optimal network for a general $N'$ is not a sub-network of the optimal network for $N > N'$ for the same robustness, $N_f$. Hence, some of the older links will need to be removed to from the optimal optimal network for $N > N'$ from the optimal network for $N'$, violating Assumption \ref{ass:as2}. We provide couple of examples to illustrate this. Consider the optimal network for $N = 10$ and $N = 12$ and robustness $N_f = 2$ (Fig. \ref{fig:opt-net-10-12-f>1}). Constructing the optimal network for $N = 12$ (bottom right of Fig. \ref{fig:opt-net-10-12-f>1}) from the optimal network for $N = 10$ (middle left of Fig. \ref{fig:opt-net-10-12-f>1}), 
will require removing two links and adding four new links. Similarly constructing the optimal network for $N = 8$ from $N = 6$ (shown in Fig. \ref{fig:net-6-8-f-1}) 
will require removing one of the links connecting the node in one half to the node in the other half. 

\begin{figure}
\subfloat{\includegraphics[width = 1.5in]{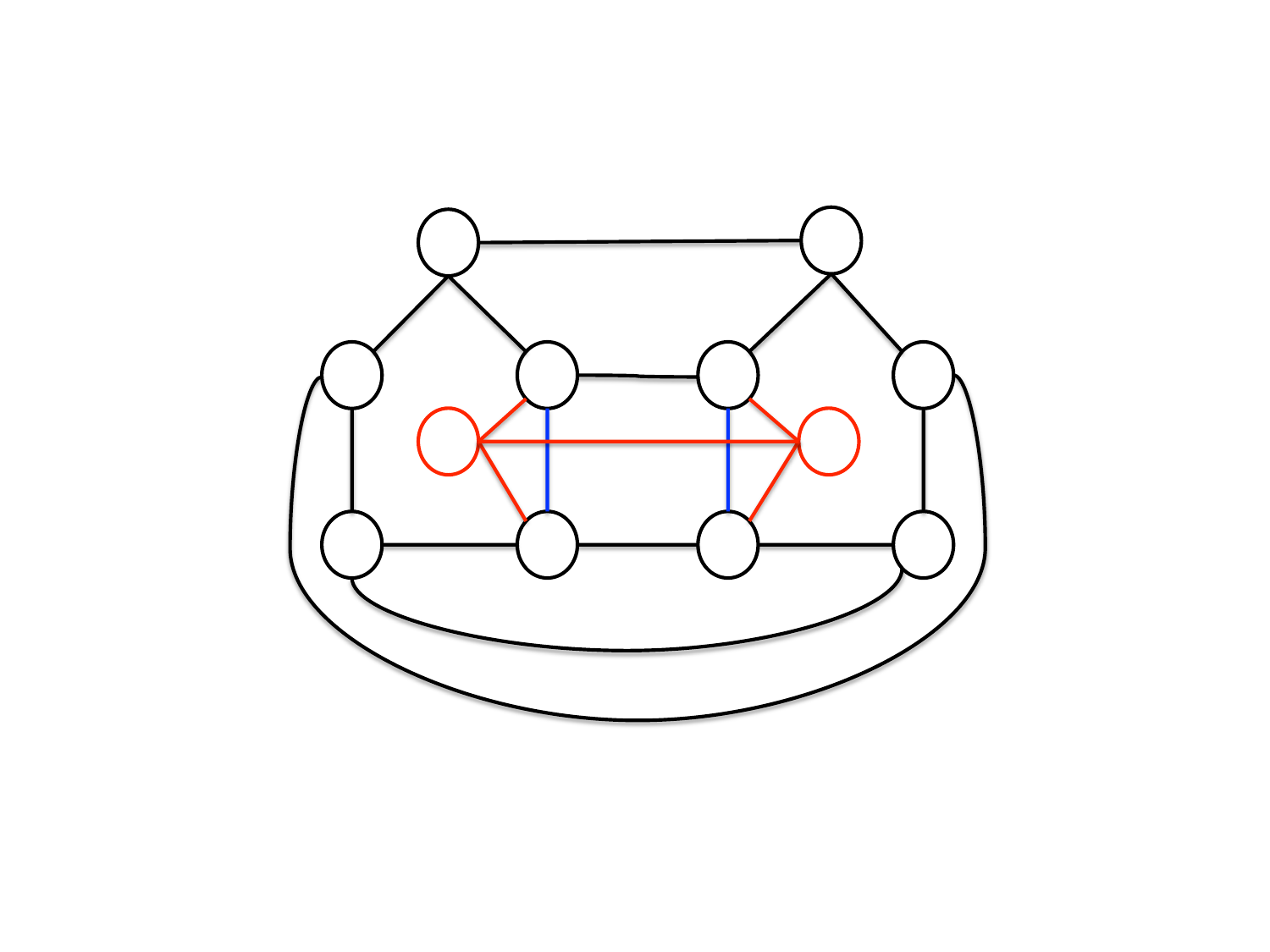}}
\subfloat{\includegraphics[width = 2in]{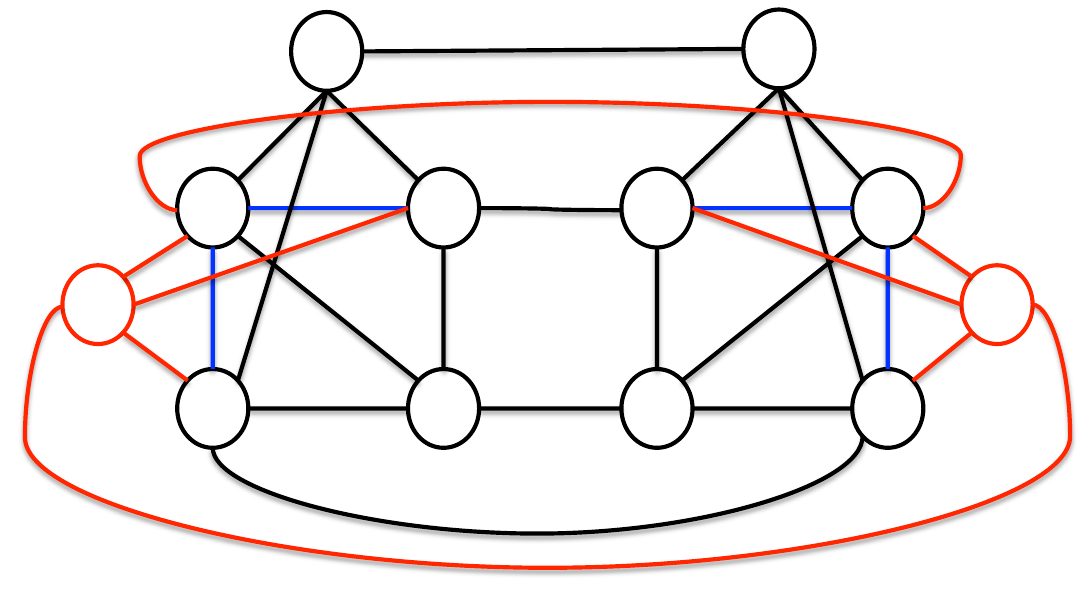}}
\caption{$12$ node network constructed from the optimal network for $N=10$. Left: $N_f = 2$. Right: $N_f = 3$.}
\label{fig:dyn-net-12-f>1} 
\end{figure}

Given this, we propose to locally construct the links around the new nodes according to the optimal network oblivious to how it modifies the connectivity of the older nodes. We provide some examples first. Consider $N_f = 2$ and $N$ changing from $10$ to $12$. Let the network when $N = 10$ be the optimal network shown in Fig. \ref{fig:opt-net-10-12-f>1}. The proposed approach connects the two new nodes according to how they would be connected in the optimal network. This is shown in Fig. \ref{fig:dyn-net-12-f>1} for $N_f = 2$ and $N_f = 3$. The new nodes and the new links are shown in `red'. Note that this myopic approach does not guarantee that the resulting network is optimal. The links that are redundant from the point of view of the optimal network are shown in `blue' in Fig. \ref{fig:dyn-net-12-f>1}. 
\begin{figure}
\begin{center}
\includegraphics[width = 3in]{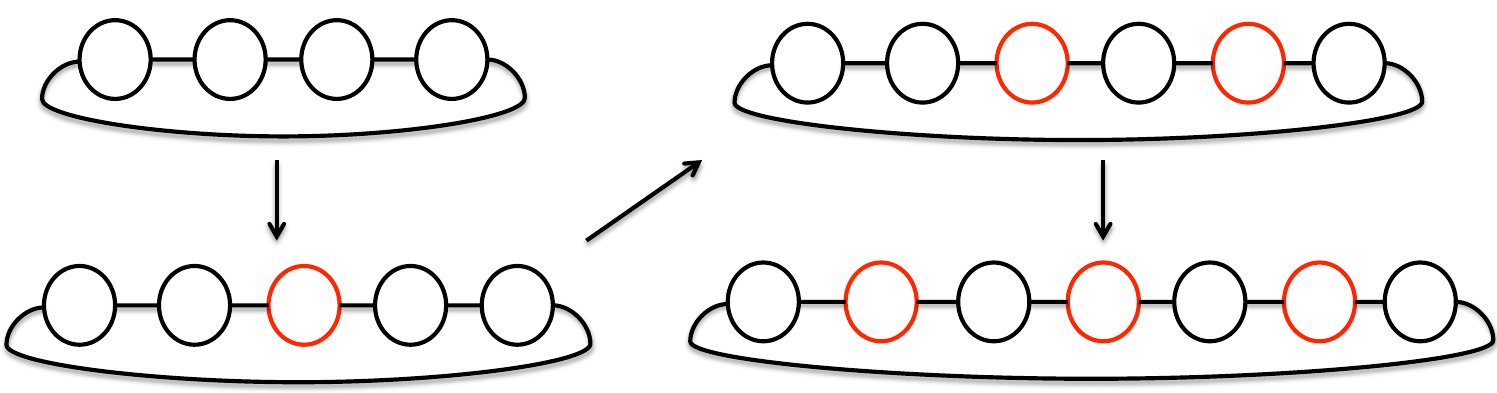}
\end{center}
\caption{Transition of $\tilde{\mathbf{G}}_2$. The new nodes are shown in red.}
\label{fig:tG2-2-tG2} 
\end{figure}

\begin{figure}
\begin{center}
\subfloat{\includegraphics[width = 1.5in]{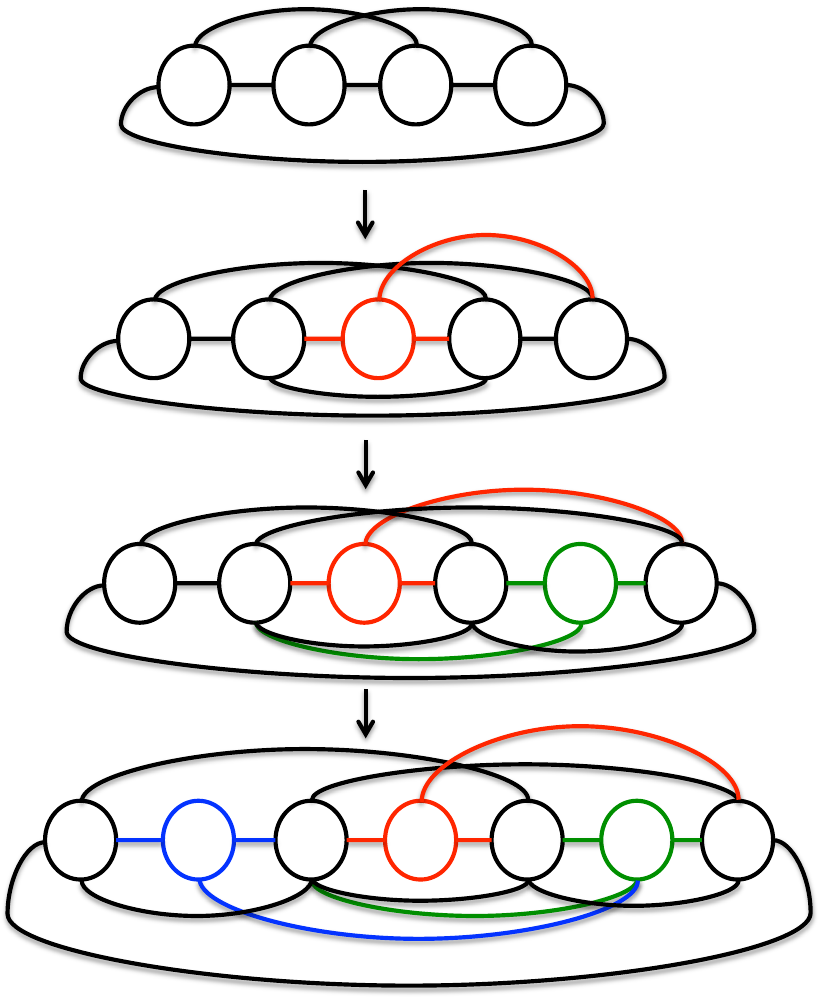}}
\subfloat{\includegraphics[width = 1.65in]{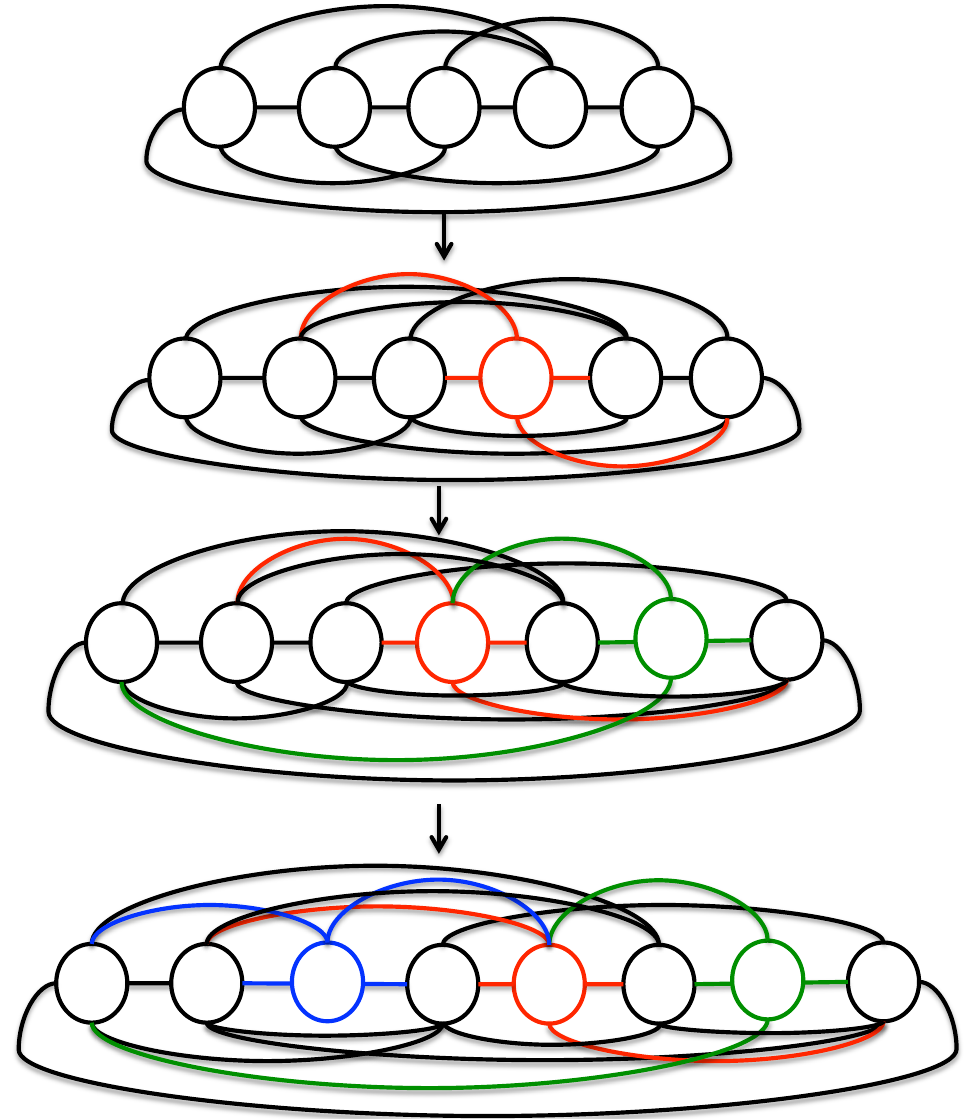}}
\end{center}
\caption{Network sequence such that each node is connected to atleast $N_f+1$ nodes, each network is a sub-network of the next network and the corresponding static network (as given by Theorem \ref{thm:even-node-connection} and Lemma \ref{lem:odd-node-connection}) is a sub-network. Left: $N_f = 2$. Right: $N_f = 3$.}
\label{fig:node-connection-sequence-2} 
\end{figure}

In the next two theorems we present the dynamic versions of the Theorem \ref{thm:even-node-connection} and Lemma \ref{lem:odd-node-connection}. We present these theorems for the dynamic setting $N_{k+1} = N_k + 1$ and for any $N_f < N_k-2 \ \forall \ k$. The two theorems present the construction of a sequence of networks such that each network in the sequence is robust and is a sub-network of the next in the sequence.  
In Fig. \ref{fig:node-connection-sequence-2} we present a couple of sequence of networks for robustness $N_f = 2$ and $N_f = 3$ respectively. In each sequence the nodes arriving at different points of time are identified by different colors. The new links corresponding to the arriving nodes are highlighted by the same color as the nodes. In each sequence the links for the arriving nodes are added according to the construction proposed in Theorem \ref{thm:even-node-connection} and Lemma \ref{lem:odd-node-connection}, ignoring how the connectivity of the older nodes are modified. We observe that the number of links of a network at any instant of time in each sequence exceeds that of the construction provided in Theorem \ref{thm:even-node-connection} and Lemma \ref{lem:odd-node-connection}. In the example on the left in Fig. \ref{fig:node-connection-sequence-2}, the second network has two additional links, the third network has three additional links, and the fourth network has four additional links. The extra links are a result of constructing the sequence such that each network is a sub-network of the next in the sequence. This constraint is necessary to satisfy Assumption \ref{ass:as2}. 

The algorithms we present in the next two theorems maintain an imaginary cycle $\tilde{\mathbf{G}}_2$ throughout that is continuously updated by adding the newly arrived nodes. We denote the $\tilde{\mathbf{G}}_2$ at time $k$ by $\tilde{\mathbf{G}}^k_2$. An example for the transition that $\tilde{\mathbf{G}}_2$ undergoes is shown in Fig. \ref{fig:tG2-2-tG2}. Here the new nodes are highlighted in `red'. We emphasize that this is only an imaginary construction that is carried out for the purpose of defining the algorithm. In the next theorem, we present the algorithm for the dynamic setting when $N_f$ is fixed and odd.
\begin{theorem}
{\it Suppose $N_k = N_{k-1}+1$ and $N_f = \tilde{N}_f-1$ is odd, consider the following procedure at time step $k$: (i) update $\tilde{\mathbf{G}}^{k-1}_{2}$ to $\tilde{\mathbf{G}}^{k}_{2}$ by inserting the new node in between any of the nodes in $\tilde{\mathbf{G}}^{k-1}_{2}$, and (ii) connect the new node to all the neighbors within $\tilde{N}_f/2$ links in $\tilde{\mathbf{G}}^{k}_{2}$. Then the resulting sequence of networks are such that $\mathbf{G}_i \subseteq \mathbf{G}_j\ \forall \ i \leq j$, $\mathbf{G}_k$ satisfies $\mathbf{RG}_{N_f} \ \forall \ k$ and $L_k \leq N_k(N_f+1)$.} 
\label{thm:sequence-even-node-connection}
\end{theorem}
Please see Appendix \ref{sec:pf-Th5} for the proof. Essentially, the idea here is to connect the new node exactly how it would be connected in the static case, i.e., according to Theorem \ref{thm:even-node-connection} or Lemma \ref{lem:odd-node-connection}, as the case maybe, ignoring how it would modify the connectivity of the older nodes. We prove the result by showing that by this approach every node is connected to its neighbors in $\tilde{\mathbf{G}}^k_2$ that are within $\tilde{N}_f/2$ links. Hence the construction provided in Theorem \ref{thm:even-node-connection} or Lemma \ref{lem:odd-node-connection}, as the case maybe, is a sub-network at any instant $k$. Consequently, the network at $k$ is $N_f-$robust or will remain connected on removing any of the $N_f$ nodes. In the next theorem, we present the algorithm for the dynamic setting when $N_f$ is fixed and even.

\begin{theorem}
{\it Suppose $N_k = N_{k-1}+1$ and $N_f = \tilde{N}_f-1$ is even, consider the following procedure at time step $k$: (i) update $\tilde{\mathbf{G}}^{k-1}_{2}$ to $\tilde{\mathbf{G}}^{k}_{2}$, (ii) connect the new node to all the neighbors within $N_f/2$ links in $\tilde{\mathbf{G}}^{k}_{2}$, (iii) if $N_k$ is even connect the new node to the diametrically opposite ($N_k/2$ links away) node in $\tilde{\mathbf{G}}^k_2$, else connect the new node to the node that is $(N_k-1)/2$ links away in $\tilde{\mathbf{G}}^k_2$ and (iv) when $N_k$ is even connect the nodes that are $N_k/2$ links apart in $\tilde{\mathbf{G}}^k_2$ if they are not already connected. Then the resulting sequence of networks are such that $\mathbf{G}_i \subseteq \mathbf{G}_j\ \forall \ i \leq j$, $\mathbf{G}_k$ satisfies $\mathbf{RG}_{N_f} \ \forall \ k$ and $L_k \leq N_k(N_f+1) + N^2_k/8$.}
\label{thm:sequence-odd-node-connection}
\end{theorem}
The idea of the update algorithm is similar to the idea in the previous theorem, that is to connect the new node as it would be connected in the static case, ignoring the changes it would effect on the other nodes. The proof style is similar to the proof of the previous theorem. Please see Appendix \ref{sec:pf-Th6} for the detailed proof. 
We note that the general case where $N_{k+1} = N_k + m_k, m_k \in \mathbb{N}$ can be analysed by assuming an imaginary time sequence indexed by $\tilde{k}$ where $N_{\tilde{k}+1} = N_{\tilde{k}} + 1$ and applying the algorithm in Theorems \ref{thm:sequence-even-node-connection} and \ref{thm:sequence-odd-node-connection} to this sequence. The network sequence for the original time sequence is then the sub-sequence given by $\{ \mathbf{G}_{\tilde{k}} \ \text{at instances when} \ N_{\tilde{k}} = N_k \}$. We note that as $N_k \rightarrow \infty$, the number of links saved is nearly $75\%$ (w.r.t the maximum $N_k(N_k-1)/2$). 

\begin{remark}
The algorithms presented in Theorems \ref{thm:sequence-even-node-connection} and \ref{thm:sequence-odd-node-connection} have complexity characterized by $\mathcal{O}(N_f)$.
\end{remark}


\subsection{When the number of nodes that can fail changes}

The idea for this setting can be motivated by the algorithm proposed in the previous section. Assume for the time being that $N_f$ is odd. In the previous section, the idea was to {\it locally connect the new node to its neighbors that are within $(N_f+1)/2$ links in the $\tilde{\mathbf{G}}_2$ that it is part of}. While this will satisfy the connectivity requirement for the new node, the older nodes may still not be connected to all of its neighbors that are $(N_f+1)/2$ links in the $\tilde{\mathbf{G}}_2$ that it is part of. This is because $N_f$ could have increased with the addition of new nodes. In that case, it is easy to see that the connectivity requirement for the older nodes can also be satisfied by adding the required links to the older nodes. 

First, we present a lemma, the equivalent of Theorems \ref{thm:sequence-even-node-connection} and \ref{thm:sequence-odd-node-connection}, that presents the approach to construct the dynamic network for the scenario where $N_f$ can change arbitrarily with time. We denote the $N_f$ at time $k$ by $N_{f,k}$, which by definition is arbitrary. Hence, this is a harder setting when compared to the setting where only a fraction of the nodes fail.
\begin{lemma}
{\it Suppose $N_k = N_{k-1}+1$. Consider the following procedure at time step $k$: (i) update $\tilde{\mathbf{G}}^{k-1}_{2}$ to $\tilde{\mathbf{G}}^{k}_{2}$, (ii) connect the new node according to Theorem \ref{thm:sequence-even-node-connection} when $N_{f,k}$ is odd, or Theorem \ref{thm:sequence-odd-node-connection} when $N_{f,k}$ is even, (iii) when $N_{f,k}$ is odd connect the older nodes to those neighbors that are within $(N_{f,k}+1)/2$ links in $\tilde{\mathbf{G}}^k_2$ and have not been already connected to, and (iv) when $N_{f,k}$ is even connect the nodes that are $N_k/2$ links away in $\tilde{\mathbf{G}}^k_2$ when $N_k$ is even and $(N_k-1)/2$ links away in $\tilde{\mathbf{G}}^k_2$ when $N_k$ is odd, if they are not already connected. Then the resulting sequence of networks are such that $\mathbf{G}_i \subseteq \mathbf{G}_j\ \forall \ i \leq j$, and $\mathbf{G}_k$ satisfies $\mathbf{RG}_{N_{f.k}} \ \forall \ k$.} 
\label{lem:sequence-dyn-node-connection}
\end{lemma}
We do not provide the proof for this lemma, since the proof is a trivial application of the proofs for Theorems \ref{thm:sequence-even-node-connection} and \ref{thm:sequence-odd-node-connection}. But unlike these theorems we cannot provide guarantee for a link bound that is any better than the number of links of the fully connected network. This is because $N_{f,k}$ is arbitrary and could equal almost the whole of the network at certain times. In the following, we consider the specific setting where only a fraction of the nodes can fail. We present algorithms for various levels of the fraction that can fail and characterize the bounds on the link usage.

In the next theorem we consider the dynamic setting with a less severe robustness requirement: $N_k = N_{k-1}+2$, $N_{f,k} = \lfloor N_k/(2f_k) \rfloor$, $f_k > 1$. The factor $1/(2f_k), (f_k > 1)$ specifies the robustness or the fraction of the nodes that can fail at any point of time without disconnecting the network. The inclusion of $1/2$ limits the robustness to utmost a $1/2$ of the number of nodes at any point of time. For this setting we can leverage the ideas developed in Theorem \ref{thm:alg-optimal-f} to present algorithms with much stricter bounds. The algorithm we are about to present maintains two imaginary $\tilde{\mathbf{G}}_{2}$s instead of one and updates them by adding the arriving nodes to the two $\tilde{\mathbf{G}}_{2}$s. 
\begin{theorem}
{\it Suppose $N_k = N_{k-1}+2$, and $N_{f,k} = \lfloor N_k/(2f_k) \rfloor$ ($f_k > 1$). Consider the following algorithm to update the network $\mathbf{G}_{k-1}$ at $k$: (i) update the $\tilde{\mathbf{G}}^{k-1}_{2}$s to the $\tilde{\mathbf{G}}^{k}_{2}$s by adding one of the new nodes to each $\tilde{\mathbf{G}}^{k-1}_{2}$, (ii) connect the two new nodes, and (iii) update the connections of the nodes in the two $\tilde{\mathbf{G}}^{k}_{2}$s according to Lemma \ref{lem:sequence-dyn-node-connection}. Then the resulting network sequence is such that $\mathbf{G}_i \subseteq \mathbf{G}_j\ \forall \ i \leq j$, $\mathbf{G}_k$ at any instant $k$ satisfies $\mathbf{RG}_{N_{f,k}}$ and $L_k \leq N^2_k/4$.}
\label{thm:seq-dyn-alg-optimal-f}
\end{theorem}

{\em Proof}: The proof for robustness is combination of the proofs in Theorems \ref{thm:sequence-even-node-connection} and \ref{thm:sequence-odd-node-connection}, and Theorem \ref{thm:alg-optimal-f}. The link count is less than $N^2_k/4$ because the nodes in a $\tilde{\mathbf{G}}^k_2$ can atmost be fully connected and there are exactly $N_k/2$ links connecting the nodes in one $\tilde{\mathbf{G}}^k_2$ to the other. 
$\blacksquare$ 

Once again, we note that the general case where $N_{k+1} = N_k + 2m_k, m_k \in \mathbb{N}$ can be analysed by assuming an imaginary time sequence indexed by $\tilde{k}$ where $N_{\tilde{k}+1} = N_{\tilde{k}} + 2$ and applying the algorithm in Theorem \ref{thm:seq-dyn-alg-optimal-f} to this sequence. 


When the number of links that can fail is further limited we can present algorithms with much improved link bound. Consider the dynamic setting: $N_{k+1} = N_k + 2m, \ m \in \mathbb{N}, \ m > 1$ and $N_{f,k} = \lfloor N_k/(2mf_k) \rfloor$, $f_k > 1$. Clearly, the robustness requirement is limited to $1/(2m)$ fraction of the total nodes at any point of time. We leverage the construction presented in Theorem \ref{thm:alg-optimal-m} for this setting. 
The algorithm we present for this setting, in contrast to the previous algorithm, maintains $2m$ cycles ($\tilde{\mathbf{G}}_{2}$s) instead of two and proceeds by updating each of the cycle as in Lemma \ref{lem:sequence-dyn-node-connection}. We present the algorithm in the theorem below.
\begin{theorem}
{\it Suppose $N_k = N_{k-1}+2m, m \in \mathbb{N}, m > 1$, and $N_{f,k} = \lfloor N_k/(2mf_k) \rfloor$, $f_k \geq 1$. Consider the following algorithm to update the network $\mathbf{G}_{k-1}$ at $k$: (i) update each of the $2m$ $\tilde{\mathbf{G}}^{k-1}_{2}$s to the $\tilde{\mathbf{G}}^{k}_{2}$s at $k$ by adding one of the $2m$ nodes to each of the $\tilde{\mathbf{G}}^{k-1}_{2}$s, (ii) connect the new nodes as a cycle, and (iii) update the connections of the nodes in each of the $2m$ $\tilde{\mathbf{G}}^{k}_{2}$s according to Lemma \ref{lem:sequence-dyn-node-connection}. Then the resulting network sequence is such that $\mathbf{G}_i \subseteq \mathbf{G}_j\ \forall \ i \leq j$, $\mathbf{G}_k$ at any instant $k$ satisfies $\mathbf{RG}_{N_{f,k}}$ and $L_k \leq N^2_k/(4m)$.}
\label{thm:seq-dyn-alg-optimal-m}
\end{theorem}
The proof is the combination of the proof of Lemma \ref{lem:sequence-dyn-node-connection} and Theorem \ref{thm:alg-optimal-m}. We note that the algorithm presented for this setting satisfies a link bound that decreases inversely with $m$. We find that as $N_k \rightarrow \infty$, the number of links that the algorithm saves is nearly $(1-1/(2m))$ fraction of the maximum number of possible links. Next, we consider the setting $f_k \leq 1$.

\subsubsection{when $f_k = 1$} 
In the next theorem we present the algorithm for robustness $N_{f,k} = N_k/2$ and characterize the bound on its link usage. 
The algorithm we present maintains two imaginary $\tilde{\mathbf{G}}_{2}$s throughout as in Theorem \ref{thm:seq-dyn-alg-optimal-f}.
\begin{theorem}
{\it Suppose $N_k = N_{k-1}+2$, and $N_{f,k} = N_k/2$. Consider the following algorithm to update the network $\mathbf{G}_{k-1}$ at $k$: (i) update each of the $\tilde{\mathbf{G}}^{k-1}_{2}$s to the $\tilde{\mathbf{G}}^{k}_{2}$s at $k$, (ii) connect the new nodes to all the other nodes in their respective $\tilde{\mathbf{G}}^{k}_{2}$s, (iii) connect the two new nodes and (iv) connect each of the new nodes to the first node in the other $\tilde{\mathbf{G}}^{k}_{2}$. Then the resulting network sequence is such that $\mathbf{G}_i \subseteq \mathbf{G}_j \ \forall i \leq j$, $\mathbf{G}_k$ satisfies $\mathbf{RG}_{N_{f,k}}$, and $\mathbf{LG}_{N_k}\ \forall \ k$.}
\label{thm:seq-dyn-alg-optimal-f1}
\end{theorem}

{\em Proof}: At any instant, the nodes in a $\tilde{\mathbf{G}}^{k}_{2}$ are fully connected. Moreover, by (iii) and (iv) of the algorithm, no set of $n < N_k/2$ in one $\tilde{\mathbf{G}}^{k}_{2}$ are connected to the same set of $n < N_k/2$ nodes in the other half. Hence, the conditions of Theorem \ref{thm:alg-optimal-f-1} are satisfied. Therefore, $\mathbf{RG}_{N_{f,k}}$ is satisfied at all times. Since the set of nodes in each of the $\tilde{\mathbf{G}}^{k}_{2}$s are fully connected, there are $N_k/4(N_k/2-1)$ links among the nodes in each of the $\tilde{\mathbf{G}}^{k}_{2}$s. In addition, by steps (iii) and (iv), there is one link connecting every node in one $\tilde{\mathbf{G}}^{k}_{2}$ to the node in the other $\tilde{\mathbf{G}}^{k}_{2}$ that arrived at the same time, and there is one additional link per node except the first two nodes. Hence, there are in total $N_k/2(N_k/2-1) + N_k/2 + N_k - 2 = N^2_k/4 + N_k - 2$ links for all $k$s. This completes the proof. $\blacksquare$

We find that as $N_k \rightarrow \infty$ the number of links that the algorithm saves is nearly $50\%$. 

\subsubsection{when $f_k < 1$}

Here we present the algorithm for the case where the number of nodes that can fail can exceed the fraction $1/2$ by a number $n$, where $n \geq 0, \ n \in \mathbb{Z}_{\geq 0}$. The robustness requirement at an instant $k$ is then given by $N_{f,k} = N_k/2+n$. In the next theorem we present an extension of the algorithm in Theorem \ref{thm:seq-dyn-alg-optimal-f1} for this case. 

\begin{theorem}
{\it Suppose $N_k = N_{k-1}+2$, and $N_{f,k} = N_k/2+n, \ n \geq 0, \ n \in \mathbb{Z}_{\geq 0}$. Consider the following algorithm to update the network $\mathbf{G}_{k-1}$ at $k$: (i) update each of the $\tilde{\mathbf{G}}^{k-1}_{2}$s to the $\tilde{\mathbf{G}}^{k}_{2}$s at $k$, (ii) connect the new nodes to all the other nodes in their respective $\tilde{\mathbf{G}}^{k}_{2}$s, (iii) connect the two new nodes and (iv) connect each of the new nodes to the first $n+1$ nodes in the other $\tilde{\mathbf{G}}^{k}_{2}$. Then the resulting network sequence is such that $\mathbf{G}_i \subseteq \mathbf{G}_j \ \forall i \leq j$, $\mathbf{G}_k$ satisfies $\mathbf{RG}_{N_{f,k}}$, and $L_k \leq N^2_k/4 +2N_k(n+3/4) -2(n+1)(n+2)$.}
\label{thm:seq-dyn-alg-optimal-fl1}
\end{theorem}
As claimed, this theorem establishes that the number of links saved by the proposed algorithm reduces atmost only linearly in $n$. Please see Appendix \ref{sec:pf-Th10} for the proof.

\begin{remark}
The algorithms in Theorems \ref{thm:seq-dyn-alg-optimal-f}, \ref{thm:seq-dyn-alg-optimal-m}, \ref{thm:seq-dyn-alg-optimal-f1} and \ref{thm:seq-dyn-alg-optimal-fl1} are characterized by complexity $\mathcal{O}(N_k)$.
\end{remark}

\begin{remark}
When nearly $1/(2m)$ fraction of the nodes can fail, by the construction
given in Theorem \ref{thm:seq-dyn-alg-optimal-f} and \ref{thm:seq-dyn-alg-optimal-f1} for $m = 1$, in Theorem  \ref{thm:seq-dyn-alg-optimal-m} for $m > 1$, and Theorem \ref{thm:seq-dyn-alg-optimal-fl1} for $m < 1$, it follows that almost all the
nodes will be connected to at most $N_k/(2m)+1$ nodes when $N_k$ is large. This by Lemma \ref{lem:minimal-edges} is almost equal to the minimum number of links required for each node when $N_k$ is large. This is the basis of our conjecture that the saving ratio achieved by our algorithms is optimal for the online setting. 
\end{remark}

\section{Conclusion}
In this work we {\it developed algorithms for designing networks that are robust to node failures and minimal in the number of links used}. We considered both the static setting and the dynamic setting where the older part of the network cannot be reformed. For the static setting, we presented {\it algorithms for constructing the optimal network in terms of the number of links used for a given node size and robustness level, the number of nodes that can fail without disconnecting the network}. For the dynamic setting, we presented online algorithms for two cases (i) when the number of nodes that can fail remains constant and (ii) when only the proportion of the nodes that can fail remains constant. We showed that for the case where $1/2$ the number of nodes can fail at any point of time, the proposed algorithm saves nearly $1/2$ of the total possible links at any point of time. In particular, we showed that when the number of nodes that can fail is limited to the fraction $1/(2m)$ ($m \in \mathbb{N}$) the proposed algorithm saves nearly as much as $(1-1/2m)$ of the total possible links at any point of time. We also showed that when the number of nodes that can fail at any point of time is $1/2$ of the number of nodes plus $n$, $n \in \mathbb{N}$, the number of links saved reduces only linearly in $n$. We conjecture that the achieved saving ratios are nearly optimal for the dynamic setting.  

\bibliographystyle{IEEEtran}
\bibliography{Refs-NetRobust}

\begin{thebibliography}{10}
\providecommand{\url}[1]{#1}
\csname url@samestyle\endcsname
\providecommand{\newblock}{\relax}
\providecommand{\bibinfo}[2]{#2}
\providecommand{\BIBentrySTDinterwordspacing}{\spaceskip=0pt\relax}
\providecommand{\BIBentryALTinterwordstretchfactor}{4}
\providecommand{\BIBentryALTinterwordspacing}{\spaceskip=\fontdimen2\font plus
\BIBentryALTinterwordstretchfactor\fontdimen3\font minus
  \fontdimen4\font\relax}
\providecommand{\BIBforeignlanguage}[2]{{%
\expandafter\ifx\csname l@#1\endcsname\relax
\typeout{** WARNING: IEEEtran.bst: No hyphenation pattern has been}%
\typeout{** loaded for the language `#1'. Using the pattern for}%
\typeout{** the default language instead.}%
\else
\language=\csname l@#1\endcsname
\fi
#2}}
\providecommand{\BIBdecl}{\relax}
\BIBdecl

\bibitem{mone2015new}
G.~Mone, ``The new smart cities,'' \emph{Communications of the ACM}, vol.~58,
  no.~7, pp. 20--21, 2015.

\bibitem{aggarwal2014evolutionary}
C.~Aggarwal and K.~Subbian, ``Evolutionary network analysis: A survey,''
  \emph{ACM Computing Surveys (CSUR)}, vol.~47, no.~1, pp. 1--36, 2014.

\bibitem{gibbons1985algorithmic}
A.~Gibbons, ``Algorithmic graph theory,'' 1985.

\bibitem{babai1995automorphism}
L.~Babai, ``Automorphism groups, isomorphism, reconstruction (chapter 27 of the
  handbook of combinatorics),'' \emph{North-Holland--Elsevier}, pp. 1447--1540,
  1995.

\bibitem{godsil2001algebraic}
C.~Godsil and G.~F. Royle, ``Algebraic graph theory,'' vol. 207, 2001.

\bibitem{dekker2004network}
A.~H. Dekker and B.~D. Colbert, ``Network robustness and graph topology,'' pp.
  359--368, 2004.

\bibitem{biggs1993algebraic}
N.~Biggs, N.~L. Biggs, and B.~Norman, ``Algebraic graph theory,'' no.~67, 1993.

\bibitem{bollobas2001random}
B.~Bollob{\'a}s and B.~B{\'e}la, ``Random graphs,'' no.~73, 2001.

\bibitem{watts1998collective}
D.~J. Watts and S.~H. Strogatz, ``Collective dynamics of
  ‘small-world’networks,'' \emph{nature}, vol. 393, no. 6684, pp. 440--442,
  1998.

\bibitem{barabasi1999emergence}
A.-L. Barab{\'a}si and R.~Albert, ``Emergence of scaling in random networks,''
  \emph{science}, vol. 286, no. 5439, pp. 509--512, 1999.

\bibitem{albert2002statistical}
R.~Albert and A.-L. Barab{\'a}si, ``Statistical mechanics of complex
  networks,'' \emph{Reviews of modern physics}, vol.~74, no.~1, p.~47, 2002.

\bibitem{barabasi2003linked}
A.-L. Barab{\'a}si, ``Linked: The new science of networks,'' 2003.

\bibitem{klemm2002highly}
K.~Klemm and V.~M. Eguiluz, ``Highly clustered scale-free networks,''
  \emph{Physical Review E}, vol.~65, no.~3, p. 036123, 2002.

\bibitem{crucitti2003efficiency}
P.~Crucitti, V.~Latora, M.~Marchiori, and A.~Rapisarda, ``Efficiency of
  scale-free networks: error and attack tolerance,'' \emph{Physica A:
  Statistical Mechanics and its Applications}, vol. 320, pp. 622--642, 2003.

\bibitem{holme2002attack}
P.~Holme, B.~J. Kim, C.~N. Yoon, and S.~K. Han, ``Attack vulnerability of
  complex networks,'' \emph{Physical review E}, vol.~65, no.~5, p. 056109,
  2002.

\bibitem{latora2004science}
V.~Latora and M.~Marchiori, ``How the science of complex networks can help
  developing strategies against terrorism,'' \emph{Chaos, solitons \&
  fractals}, vol.~20, no.~1, pp. 69--75, 2004.

\bibitem{criado2006new}
R.~Criado, A.~G. del Amo, B.~Hern{\'a}ndez-Bermejo, and M.~Romance, ``New
  results on computable efficiency and its stability for complex networks,''
  \emph{Journal of Computational and Applied Mathematics}, vol. 192, no.~1, pp.
  59--74, 2006.

\bibitem{criado2005effective}
R.~Criado, J.~Flores, B.~Hern{\'a}ndez-Bermejo, J.~Pello, and M.~Romance,
  ``Effective measurement of network vulnerability under random and intentional
  attacks,'' \emph{Journal of Mathematical Modelling and Algorithms}, vol.~4,
  no.~3, pp. 307--316, 2005.

\bibitem{dekker2005symmetry}
A.~H. Dekker and B.~Colbert, ``The symmetry ratio of a network,'' pp. 13--20,
  2005.

\bibitem{manzano2013endurance}
M.~Manzano, E.~Calle, V.~Torres-Padrosa, J.~Segovia, and D.~Harle, ``Endurance:
  A new robustness measure for complex networks under multiple failure
  scenarios,'' \emph{Computer Networks}, vol.~57, no.~17, pp. 3641--3653, 2013.

\bibitem{piraveenan2013quantifying}
M.~Piraveenan, G.~Thedchanamoorthy, S.~Uddin, and K.~S.~K. Chung, ``Quantifying
  topological robustness of networks under sustained targeted attacks,''
  \emph{Social Network Analysis and Mining}, vol.~3, no.~4, pp. 939--952, 2013.

\bibitem{schieber2016information}
T.~A. Schieber, L.~Carpi, A.~C. Frery, O.~A. Rosso, P.~M. Pardalos, and M.~G.
  Ravetti, ``Information theory perspective on network robustness,''
  \emph{Physics Letters A}, vol. 380, no.~3, pp. 359--364, 2016.

\bibitem{bigdeli2009comparison}
A.~Bigdeli, A.~Tizghadam, and A.~Leon-Garcia, ``Comparison of network
  criticality, algebraic connectivity, and other graph metrics,'' pp. 1--6,
  2009.

\bibitem{liu2017comparative}
J.~Liu, M.~Zhou, S.~Wang, and P.~Liu, ``A comparative study of network
  robustness measures,'' \emph{Frontiers of Computer Science}, vol.~11, no.~4,
  pp. 568--584, 2017.

\bibitem{bilal2013characterization}
K.~Bilal, M.~Manzano, S.~U. Khan, E.~Calle, K.~Li, and A.~Y. Zomaya, ``On the
  characterization of the structural robustness of data center networks,''
  \emph{IEEE Transactions on Cloud Computing}, vol.~1, no.~1, pp. 1--1, 2013.

\bibitem{manzano2013connectivity}
M.~Manzano, K.~Bilal, E.~Calle, and S.~U. Khan, ``On the connectivity of data
  center networks,'' \emph{IEEE Communications Letters}, vol.~17, no.~11, pp.
  2172--2175, 2013.

\bibitem{tizghadam2009autonomic}
A.~Tizghadam and A.~Leon-Garcia, ``Autonomic traffic engineering for network
  robustness,'' \emph{IEEE journal on selected areas in communications},
  vol.~28, no.~1, pp. 39--50, 2009.

\bibitem{tizghadam2008robust}
------, ``On robust traffic engineering in transport networks,'' pp. 1--6,
  2008.

\bibitem{rueda2017robustness}
D.~F. Rueda, E.~Calle, and J.~L. Marzo, ``Robustness comparison of 15 real
  telecommunication networks: Structural and centrality measurements,''
  \emph{Journal of Network and Systems Management}, vol.~25, no.~2, pp.
  269--289, 2017.

\bibitem{chen2017robustness}
Z.~Chen, J.~Wu, Y.~Xia, and X.~Zhang, ``Robustness of interdependent power
  grids and communication networks: A complex network perspective,'' \emph{IEEE
  Transactions on Circuits and Systems II: Express Briefs}, vol.~65, no.~1, pp.
  115--119, 2017.

\bibitem{wang2017multi}
X.~Wang, Y.~Ko{\c{c}}, S.~Derrible, S.~N. Ahmad, W.~J. Pino, and R.~E. Kooij,
  ``Multi-criteria robustness analysis of metro networks,'' \emph{Physica A:
  Statistical Mechanics and its Applications}, vol. 474, pp. 19--31, 2017.

\bibitem{lordan2016robustness}
O.~Lordan, J.~M. Sallan, N.~Escorihuela, and D.~Gonzalez-Prieto, ``Robustness
  of airline route networks,'' \emph{Physica A: Statistical Mechanics and its
  Applications}, vol. 445, pp. 18--26, 2016.

\bibitem{zhou2019efficiency}
Y.~Zhou, J.~Wang, and G.~Q. Huang, ``Efficiency and robustness of weighted air
  transport networks,'' \emph{Transportation Research Part E: Logistics and
  Transportation Review}, vol. 122, pp. 14--26, 2019.

\bibitem{pallis2009structure}
G.~Pallis, D.~Katsaros, M.~D. Dikaiakos, N.~Loulloudes, and L.~Tassiulas, ``On
  the structure and evolution of vehicular networks,'' pp. 1--10, 2009.

\bibitem{angskun2007binomial}
T.~Angskun, G.~Bosilca, and J.~Dongarra, ``Binomial graph: A scalable and
  fault-tolerant logical network topology,'' pp. 471--482, 2007.

\bibitem{yaziciouglu2015formation}
A.~Y. Yaz{\i}c{\i}o{\u{g}}lu, M.~Egerstedt, and J.~S. Shamma, ``Formation of
  robust multi-agent networks through self-organizing random regular graphs,''
  \emph{IEEE Transactions on Network Science and Engineering}, vol.~2, no.~4,
  pp. 139--151, 2015.

\bibitem{pedersen2006applying}
J.~M. Pedersen, A.~Patel, T.~P. Knudsen, and O.~B. Madsen, ``Applying 4-regular
  grid structures in large-scale access networks,'' \emph{Computer
  communications}, vol.~29, no.~9, pp. 1350--1362, 2006.

\bibitem{bellingeri2018robustness}
M.~Bellingeri and D.~Cassi, ``Robustness of weighted networks,'' \emph{Physica
  A: Statistical Mechanics and its Applications}, vol. 489, pp. 47--55, 2018.

\bibitem{ramanathan2000topology}
R.~Ramanathan and R.~Rosales-Hain, ``Topology control of multihop wireless
  networks using transmit power adjustment,'' vol.~2, pp. 404--413, 2000.

\bibitem{borbash2002distributed}
S.~A. Borbash and E.~H. Jennings, ``Distributed topology control algorithm for
  multihop wireless networks,'' vol.~1, pp. 355--360, 2002.

\bibitem{li2004topology}
N.~Li and J.~C. Hou, ``Topology control in heterogeneous wireless networks:
  Problems and solutions,'' vol.~1, 2004.

\bibitem{venuturumilli2010obtaining}
A.~Venuturumilli and A.~Minai, ``Obtaining robust wireless sensor networks
  through self-organization of heterogeneous connectivity,'' pp. 487--494,
  2010.

\bibitem{qiu2017rose}
T.~Qiu, A.~Zhao, F.~Xia, W.~Si, and D.~O. Wu, ``Rose: Robustness strategy for
  scale-free wireless sensor networks,'' \emph{IEEE/ACM Transactions on
  Networking}, vol.~25, no.~5, pp. 2944--2959, 2017.

\end{thebibliography}

\begin{appendices}

\section{Proof of Theorem \ref{thm:even-node-connection}}
\label{sec:pf-Th1}

We start by showing that we can construct the network according to the steps in the theorem. {\em Case, $\tilde{N}_f = 2$}: Index the nodes arbitrarily. Connect the first node to second, second to third, third to fourth and so on and connect the $N$th node back to the first node. In such a construction each node is connected to two other nodes. Call this network $\tilde{\mathbf{G}}_2$. This proves this case. {\em Case, $\tilde{N}_f = k$ is even}: We prove once again by giving a construction. Connect every node as in case $\tilde{N}_f = 2$ to form $\tilde{\mathbf{G}}_2$. Connect every node to its neighbor that are $2$, $3$,...,$k/2$ links apart (there are two neighbors of the same distance) in $\tilde{\mathbf{G}}_2$. 
Call this network $\tilde{\mathbf{G}}_k$. {\em Case, $\tilde{N}_f = k$ is odd}: Start with the network for $k-1$ (which is even) constructed as per the previous case. Call this network $\tilde{\mathbf{G}}_{k-1}$. Connect the first node to the node that is $N/2$ links apart in $\tilde{\mathbf{G}}_2$. Then the second node with the node that is $N/2$ links apart as per $\tilde{\mathbf{G}}_2$ if it has not already been connected and so on. Since $N$ is even there is a unique pair of nodes that are $N/2$ links apart for every node in $\tilde{\mathbf{G}}_2$. 
This covers all the cases and proves that we can construct the network according to the steps in the theorem.

To prove robustness, first we consider the case where $N_f = N-2$. In this case $\tilde{N}_f = N-1$ and so by construction every node will be connected to every other node. This network is fully connected and so will satisfy robustness $N_f = N-2$. Next, we consider the case $N_f \leq N-3$. Can a set of $k \leq \lfloor (N - N_1 )/2 \rfloor$ ($k > 2$) nodes be isolated suppose $N_1 \leq N_f$ nodes are removed from the network constructed as per the theorem? 
We consider the extreme case where most of all the links from the $k$ nodes are connected to the $k$ nodes themselves and the $N_1$ nodes that were removed. We consider the two cases when $\tilde{N}_f$ is even and odd separately.

{\it $\tilde{N}_f$ is even}: For the construction in the Theorem this happens when the $k$ nodes and the $N_1$ nodes are the nodes that are side by side as per $\tilde{\mathbf{G}}_2$ (see above for the definition of $\tilde{\mathbf{G}}_{2}$). In this case there cannot be more than $\lfloor N_1/2 \rfloor \leq \lfloor N_f/2 \rfloor$ of the $N_1$ nodes on the extremes of both the sides (as per $\tilde{\mathbf{G}}_{2}$) of the set of $k$ and $N_1$ nodes. Then the extreme node of the $k$ nodes towards the side (as per $\tilde{\mathbf{G}}_2$) that has less than $\lfloor N_1/2 \rfloor$ of the $N_1$ nodes will have atmost $\lfloor N_f/2 \rfloor$ of the $N_1$ nodes as its neighbors on the same side (as per $\tilde{\mathbf{G}}_2$). This is less than $\tilde{N}_f/2$ of the neighbors it is connected to on one side (as per $\tilde{\mathbf{G}}_2$) by atleast one. So this extreme node will be connected to atleast one node of the other $N/2-k-N_1$ nodes.

{\it $\tilde{N}_f$ is odd}: Given that $k > 2$, $N_1 \leq N_f < N - 4$, i.e., $\tilde{N}_f < N - 3$. Similar to earlier, there cannot be more than $N_f/2$ of the $N_1$ nodes on the extremes of both the sides of the $k$ and $N_1$ nodes. Note that, by construction each node is connected to its neighbors that are within $N_f/2$ links on its either side as per $\tilde{\mathbf{G}}_2$. Hence, when there can be atmost $N_f/2-1$ of the $N_1$ nodes on the extreme of one of the sides (as per $\tilde{\mathbf{G}}_2$) of the $k$ and $N_1$ nodes, by the same argument as in the `{\it $\tilde{N}_f$ is even}' case, it follows that the extreme node of the $k$ nodes towards this side will be connected to atleast one node of the other $N-k-N_1$ nodes. The only remaining scenario to consider here is when $N_1 = N_f$ and there are exactly $N_f/2$ nodes on the extreme of one side of the $k$ and $N_1$ nodes (as per $\tilde{\mathbf{G}}_2$). In this case there can be $N_f/2$ of the $N_1$ nodes on the extreme of the other (second) side (as per $\tilde{\mathbf{G}}_2$). The number of nodes on either side of one of the nodes (denote by n1) of the $k$ nodes (there are atleast 3 of them) between the two nodes (of the $k$ nodes) that are towards the extreme of both the sides (as per $\tilde{\mathbf{G}}_2$ of the $k$ and $N_1$ nodes) will be $\leq k-2+N_f/2 = k+\sup\{N_1\}/2-2 \leq N/2-2 < N/2-1$. This then implies that the node that is $N/2$ links apart from the n1 node will be part of the remaining $N-k-N_1$ nodes of its half. The construction is such that the nodes that are $N/2$ links apart as per $\tilde{\mathbf{G}}_2$ are connected. So, in this case, there is always a node of the $k$ nodes that will remain connected to one of the other $N-k-N_1$ of nodes. 

The case considered above by definition is the worst possible scenario and so it follows that no set of $k \leq \lfloor (N - N_1 )/2 \rfloor$ nodes can be isolated when $N_1 \leq N_f$ nodes are removed. This in turn implies that no set of $k < N-N_1$ can be isolated. This implies the network will remain connected after removing any of $N_1 \leq N_f$ nodes. In addition, by construction each node is connected to just $\tilde{N}_f = N_f+1$ nodes, which by Lemma \ref{lem:minimal-edges} is the minimal that is required. Hence the network is optimal $\blacksquare$

\section{Proof of Lemma \ref{lem:static-rob-f-1}} 
\label{sec:pf-Lm2}

For $N = 8$ it is necessary that every column of $\bf{L}$ has no more than $3$ zeros. If not there is guaranteed to be a $4$ node sub network that is not connected. This implies that it is necessary that $L \geq L_{opt}(8)$. For a network with $N$ nodes it is necessary that every column of the corresponding matrix $\bf{L}$ has at most $\frac{N}{2} - 1$ zeros. If not there is guaranteed to be a $\frac{N}{2}$ node sub network that is not connected. Hence it is necessary that $L \geq L^1_{opt}(N)$. This proves the first part in general.

For the second part we start by giving a construction of $\bf{G}$ for different values of $m$ that has $L^1_{opt}(N)$ links and show that this satisfies the robustness property. The construction of $\mathbf{G}$ for $N = 8$, $N = 10, N = 12, ...$ is shown below. Let the network in this sequence for a general $N$ be $\mathbf{G}^{opt}_N$ and the corresponding adjacency matrix be $\bf{L}^{1,opt}_N$.
\[
\bf{L}^{1,opt}_8 = \left[\begin{array} {c|c|c|c|c|c|c|c}
0 & 1 & 1 & 1 & 0 & 1 & 0 & 1 \\
\hline
1 & 0 & 1 & 1 & 1 & 0 & 1 & 0 \\
\hline
1 & 1 & 0 & 1 & 1 & 1 & 0 & 0 \\
\hline
1 & 1 & 1 & 0 & 0 & 0 & 1 & 1 \\
\hline
0 & 1 & 1 & 0 & 0 & 1 & 1 & 1 \\
\hline
1 & 0 & 1 & 0 & 1 & 0 & 1 & 1 \\
\hline
0 & 1 & 0 & 1 & 1 & 1 & 0 & 1 \\
\hline
1 & 0 & 0 & 1 & 1 & 1 & 1 & 0 \\
\end{array} \right],\] \[ \bf{L}^{1,opt}_{10} = \left[\begin{array} {c|c|c|c|c|c|c|c|c|c}
0 & 1 & 1 & 1 & 1 & \color{blue} 0 & \color{blue} 0 &\color{red} 1 &\color{blue} 0 & 1 \\
\hline
1 & 0 & 1 & 1 & 1 & \color{blue} 0 & \color{red}1 &\color{blue} 0 & 1 &\color{blue} 0 \\
\hline
1 & 1 & 0 & 1 & 1 & 1 &\color{blue} 0 & 1 &\color{blue} 0 &\color{blue} 0 \\
\hline
1 & 1 & 1 & 0 & 1 & 1 & 1 &\color{blue} 0 &\color{blue} 0 &\color{blue} 0 \\
\hline
1 & 1 & 1 & 1 &\color{blue} 0 & \color{blue} 0 & \color{blue} 0 & \color{blue} 0 & 1 & 1 \\
\hline
\color{blue} 0 & \color{blue} 0 & 1 & 1 & \color{blue} 0 & \color{blue} 0 & 1 & 1 & 1 & 1 \\
\hline
\color{blue} 0 & \color{red} 1 & \color{blue} 0 & 1 &\color{blue} 0 & 1 & 0 & 1 & 1 & 1 \\
\hline
\color{red}1 & \color{blue} 0 & 1 & \color{blue} 0 & \color{blue} 0 & 1 & 1 & 0 & 1 & 1 \\
\hline
\color{blue} 0 & 1 & \color{blue} 0 & \color{blue} 0 & 1 & 1 & 1 & 1 & 0 & 1 \\
\hline
1 & \color{blue} 0 & \color{blue} 0 & \color{blue} 0 & 1 & 1 & 1 & 1 & 1 & 0
\end{array} \right], \]\[ \bf{L}^{1,opt}_{12}  = \left[\begin{array} {c|c|c|c|c|c|c|c|c|c|c|c}
0 & 1 & 1 & 1 & 1 & 1 & \color{blue} 0 &\color{blue} 0 & \color{blue}0 & \color{red} 1 & \color{blue} 0 & 1 \\
\hline
1 & 0 & 1 & 1 & 1 & 1 & \color{blue} 0 & \color{blue} 0 &\color{red} 1 & \color{blue}0 & 1 &\color{blue} 0 \\
\hline
1 & 1 & 0 & 1 & 1 & 1 & \color{blue} 0 &\color{red} 1 & \color{blue} 0 & 1 &\color{blue} 0 &\color{blue} 0 \\
\hline
1 & 1 & 1 & 0 & 1 & 1 & 1 & \color{blue} 0 & 1 &\color{blue} 0 & \color{blue} 0 &\color{blue} 0 \\
\hline
1 & 1 & 1 & 1 & 0 & 1 & 1 & 1 & \color{blue} 0 & \color{blue} 0 & \color{blue} 0 &\color{blue} 0 \\
\hline
1 & 1 & 1 & 1 & 1 & \color{blue} 0 & \color{blue} 0 & \color{blue} 0 &\color{blue} 0 &\color{blue} 0 & 1 & 1 \\
\hline
\color{blue} 0 & \color{blue} 0 & \color{blue} 0 & 1 & 1 & \color{blue} 0 & \color{blue} 0 & 1 & 1 & 1 & 1 & 1 \\
\hline
\color{blue} 0 & \color{blue} 0 & \color{red} 1 & \color{blue} 0 & 1 & \color{blue} 0 & 1 & 0 & 1 & 1 & 1 & 1 \\
\hline
\color{blue} 0 & \color{red}1 & \color{blue} 0 & 1 & \color{blue} 0 & \color{blue} 0 & 1 & 1 & 0 & 1 & 1 & 1 \\
\hline
\color{red} 1 & \color{blue} 0 & 1 & \color{blue} 0 & \color{blue} 0 &\color{blue} 0 & 1 & 1 & 1 & 0  & 1 & 1 \\
\hline
\color{blue} 0 & 1 & \color{blue} 0 & \color{blue} 0 & \color{blue}0 & 1 & 1 & 1 & 1 & 1  & 0 & 1 \\
\hline
1 & \color{blue} 0 & \color{blue} 0 &\color{blue} 0 &\color{blue} 0 & 1 & 1 & 1 & 1 & 1  & 1 & 0 \\
\end{array} \right], ...\]
The construction given above satisfies the following: {\it there are two fully connected halves}, (ii) {\it each node in each half is connected to two other nodes in the other half such that no set of $n < N/2$ nodes in one half are connected to the same set of $n < N/2$ nodes in the other half}. Given this the steps in the proof of Theorem \ref{thm:alg-optimal-f-1} can be followed to prove the second part $\blacksquare$

\section{Proof of Theorem \ref{thm:alg-optimal-f-1}} 
\label{sec:pf-Th2}

First, we observe that the number of links by this construction is equal to $L^1_{opt}(N)$. So the algorithm constructs a network with the optimal number of links. Third, the construction $\bf{L}^{opt}_N$ in Lemma \ref{lem:static-rob-f-1} is an example a network that satisfies the algorithm. Hence, such a network exists. Next, we show that the robustness property is satisfied by the algorithm. First, no single node can be isolated by removing any of the $N/2$ nodes because by this construction each node has $N/2+1$ links. 
Can a set of two nodes from a half be isolated by removing any of the $N/2$ of the nodes? First we consider the case where these two new nodes are from the same half. Then it has to be that $N/2-2$ of the nodes of the half they belong to are removed and only two from the other half are removed. By construction these two nodes are connected. 
And, since by construction no two nodes from the same half are connected to the same two nodes in the other half there is one other node in the other half that these two nodes will remain connected to. 
This completes this part. 

Similarly can a set $n < N/2$ nodes from a half be isolated by removing a $N/2$ node sub-network. Then it has to be that $N/2-n$ are removed from the same half these nodes belong to and that $n$ nodes are removed from the other half. By construction these $n$ nodes will remain connected. And, since by construction no set of $n$ nodes are connected to the same set of $n$ nodes in the other half there is one other node in the other half that the $n$ nodes in point will remain connected to. Thus no set of $n$ nodes can be isolated. The remaining case is removing $N/2$ nodes from the same half. In this case one half is completely removed and the remaining nodes are from the other half which by construction is fully connected. This completes the proof $\blacksquare$

\section{Proof of Lemma \ref{lem:odd-node-connection}} 
\label{sec:pf-Lm3}

We start by giving a construction to show that we can construct a network according to the steps given in the lemma. The steps for the case $N_f = 2$ and $\tilde{N}_f = k$ when $k$ is even are the same as in the proof for Theorem \ref{thm:even-node-connection}. {\em Case, $\tilde{N}_f = k$ is odd}: Start with the network $\tilde{\mathbf{G}}_{k-1}$ (see Theorem \ref{thm:even-node-connection} for definition of $\tilde{\mathbf{G}}_{k-1}$). In this case, for every node there are two nodes that are $(N-1)/2$ links apart as per $\tilde{\mathbf{G}}_{2}$. Connect the first node to the node that is $(N-1)/2$ links apart towards one side as per $\tilde{\mathbf{G}}_{2}$, the second node to the node that is $(N-1)/2$ links apart towards the same side as per $\tilde{\mathbf{G}}_{2}$ and so on. This results in a network where only the node labelled by $1$ is connected to $\tilde{N}_f+1$ nodes, and the others are connected to $\tilde{N}_f$ nodes. Since the indexing of the nodes is arbitrary we could have arranged the nodes in any sequence. 
This proves that we can construct the network according the steps in the lemma.

Next, we prove robustness. First we consider the case where $N_f = N-2$. In this case $\tilde{N}_f = N-1$ and so by construction every node will be connected to every other node. This network is fully connected and so will satisfy robustness $N_f = N-2$. Next, we consider the case $N_f \leq N-3$. Can a set of $k \leq \lfloor (N - N_1 )/2 \rfloor$ ($k > 2$) nodes be isolated suppose $N_1 \leq N_f$ nodes are removed from the network constructed as per the theorem? We consider the extreme case where most of all the links from the $k$ nodes are connected to the $k$ nodes themselves and the $N_1$ nodes that were removed and then show separately that $N_f$-robustness holds when  $\tilde{N}_f$ is even and odd. For the construction in the Theorem this happens when the $k$ nodes and the $N_1$ nodes are the nodes that are side by side as per $\tilde{\mathbf{G}}_2$. 

The remaining steps to prove are same as the proof for Theorem \ref{thm:even-node-connection}. Hence we do not repeat the steps here. The only scenario that is different is when $\tilde{N}_f$ is odd, $N_1 = N_f$ and there are exactly $N_f/2$ nodes on the extreme of one side of the $k$ and $N_1$ nodes (as per $\tilde{\mathbf{G}}_2$). In this case the number of nodes on either side (as per $\tilde{\mathbf{G}}_2$) of one of the nodes (denote by n1) of the $k$ nodes (there are atleast 3 of them) between the two nodes (of the $k$ nodes) that are towards the extreme of both the sides (as per $\tilde{\mathbf{G}}_2$ of the $k$ and $N_1$ nodes) will be $\leq k-2+N_f/2 = k+\sup\{N_1\}/2-2 \leq \lfloor N/2 \rfloor -2 < \lfloor N/2 \rfloor -1$. This then implies that both the nodes that are $(N-1)/2$ links apart from the n1 node will be part of the remaining $N-k-N_1$ nodes of its half. The construction by Af for this case is such that every node is connected to atleast one node that is $(N-1)/2$ links apart as per $\tilde{\mathbf{G}}_2$. Combining all the observations we can conclude that there is always a node of the $k$ nodes that will remain connected to one of the other $N/2-k-N_1$ of nodes. Then following the same steps as in the proof of Theorem \ref{thm:even-node-connection} it follows that $N_f$-robustness is satisfied $\blacksquare$

\section{Proof of Theorem \ref{thm:alg-optimal-f}} 
\label{sec:pf-Th3}

First, when $N_f$ nodes are removed completely from one half, then the whole network remains connected because the other half by construction is connected and each of the remaining nodes in the first half are connected to one node in the other half. By construction and Theorem \ref{thm:even-node-connection} and Lemma \ref{lem:odd-node-connection} the two halves are by themselves robust to failure of $N_f-1$ nodes in their respective halves. Hence, when $N_1 < N_f$ are removed from one half and $N_f-N_1$ are removed from the other half, each of the halves remain connected. Moreover, the construction is such that there is a node from one half that is connected to a node in the other half after removing any of the $N_1$ nodes. Hence, the whole network remains connected. 
All of the above observations together imply that the network will remain connected after removing any of the $N_f$ nodes. This implies that the network constructed by Af satisfies robustness for a given $f$. Moreover, each node is connected to $N_f+1$ nodes by construction, which by Lemma \ref{lem:minimal-edges} is the minimal number of connections that is required for robustness to hold for a given $f$. Hence, the construction by Af for a given $f$ is optimal $\blacksquare$

\section{Proof of Theorem \ref{thm:alg-optimal-m}}
\label{sec:pf-Th4}

We prove this case by case. Lets consider the case where $N/(2m)-k (0 < k < N/(2m))$ nodes are removed from one set. The remaining $k$ nodes in this set are connected by construction. Now by construction to isolate this set of $k$ nodes it is necessary that the two nodes they are connected to in their respective cycles are removed. By construction these nodes are unique. Hence $2k$ more nodes would have to be removed to isolate the $k$ nodes in point. Thus the total number of nodes needed to be removed to isolate the $k$ nodes in point exceeds $N/(2m)$ by $k > 0$. Thus no set of $k < N/(2m)$ nodes from a set can be isolated after removing any of the other $N/2m$ nodes. 

The only remaining case to consider is when $N/(2m)$ nodes are removed from one of the $2m$ sets. In this case the remaining nodes are connected because (i) each of the remaining $2m-1$ sets are fully connected by construction, and (ii) each node in each of the remaining set is connected to atleast one other node from one of the other sets in their respective cycles $\blacksquare$

\section{Proof of Theorem \ref{thm:sequence-even-node-connection}} 
\label{sec:pf-Th5}

We prove robustness by induction. Lets assume that nodes are connected to their neighbors $\tilde{N}_f/2$ links away in $\tilde{\mathbf{G}}^{k-1}_{2}$ at $k-1$. At time instant $k$ the new node is added between two nodes of $\tilde{\mathbf{G}}^{k-1}_{2}$ to form $\tilde{\mathbf{G}}^{k}_{2}$. This shifts some of the neighbors that were within $\tilde{N}_f/2$ links in $\tilde{\mathbf{G}}^{k-1}_{2}$ for some of the nodes by one location. By construction of $\tilde{\mathbf{G}}^{k}_{2}$, these nodes should be within $\tilde{N}_f/2$ links from the new node in $\tilde{\mathbf{G}}^{k}_{2}$. Then given that the new node is connected to the nodes that are within $\tilde{N}_f/2$ links in $\tilde{\mathbf{G}}^{k}_{2}$, each of those nodes should be connected in the network $\mathbf{G}_k$ to its neighbors that are within $\tilde{N}_f/2$ links in $\tilde{\mathbf{G}}^{k}_{2}$. Thus all the nodes in $\mathbf{G}_k$ are connected to their neighbors that are within $\tilde{N}_f/2$ links in $\tilde{\mathbf{G}}^{k}_{2}$. The connectivity is trivially valid at time $k=1$ because there is no history before $k = 1$ and Theorem \ref{thm:even-node-connection} or Lemma \ref{lem:odd-node-connection} can be directly applied, whichever is applicable, to construct the network at $k = 1$. Then by induction it follows that the nodes are connected to their neighbors $\tilde{N}_f/2$ links away in $\tilde{\mathbf{G}}^{k}_{2}$ for all $k$s. The resulting sequence of networks $\{\mathbf{G}_k\}$ is such that the network in Theorem \ref{thm:even-node-connection} for $N = N_k$ when $N_k$ is even, and in Lemma \ref{lem:odd-node-connection} for $N = N_k$ when $N_k$ is odd is a sub-network of $\mathbf{G}_k$. By Theorem \ref{thm:even-node-connection} and Lemma \ref{lem:odd-node-connection} it then follows that $\mathbf{G}_k$ satisfies $N_f$-robustness for all $k$s. 

The networks by construction satisfy $\mathbf{G}_i \subseteq \mathbf{G}_j$ for all $i \leq j$. On arrival each new node is connected to $\tilde{N}_f$ other nodes within the same $\tilde{\mathbf{G}}^k_2$. Hence, at every step $\tilde{N}_f$ links are formed for each of the new nodes to connect it to the other nodes in the respective $\tilde{\mathbf{G}}^k_2$s. This corresponds to a total of $N_k\tilde{N}_f$ links by time step $k$. This completes the proof $\blacksquare$

\section{Proof of Theorem  \ref{thm:sequence-odd-node-connection}}
\label{sec:pf-Th6}

We prove robustness by induction. Lets assume that nodes are connected to their neighbors $N_f/2$ links away in $\tilde{\mathbf{G}}^{k-1}_{2}$ at $k-1$. At time instant $k$ the new node is added between two nodes of $\tilde{\mathbf{G}}^{k-1}_{2}$ to form $\tilde{\mathbf{G}}^{k}_{2}$. This shifts some of the neighbors that were within $N_f/2$ links in $\tilde{\mathbf{G}}^{k-1}_{2}$ for some of the nodes by one location. By construction of $\tilde{\mathbf{G}}^{k}_{2}$, these nodes should be within $N_f/2$ links from the new node in $\tilde{\mathbf{G}}^{k}_{2}$. Then given that the new node is connected to the nodes that are within $N_f/2$ links in $\tilde{\mathbf{G}}^{k}_{2}$, each of those nodes should be connected in the network $\mathbf{G}_k$ to its neighbors that are within $N_f/2$ links in $\tilde{\mathbf{G}}^{k}_{2}$. Thus all the nodes in $\mathbf{G}_k$ are connected to their neighbors that are within $N_f/2$ links in $\tilde{\mathbf{G}}^{k}_{2}$.

Let $N_{k-1}$ be even. Note that by construction each node in $\mathbf{G}_{k-1}$ is connected to the node that is diametrically opposite ($N_{k-1}/2$ links away) in $\tilde{\mathbf{G}}^{k-1}_2$. Then at $k$, updating $\tilde{\mathbf{G}}^{k-1}_2$ to $\tilde{\mathbf{G}}^{k}_2$ as in the statement makes the diametrically opposite nodes (which are $N_{k-1}/2$ links apart) in $\tilde{\mathbf{G}}^{k-1}_2$, $(N_k-1)/2$ links apart in $\tilde{\mathbf{G}}^{k}_2$. These pair of nodes are connected by assumption. 
In addition by construction the new node will be connected to one of the nodes that is $(N_k-1)/2$ links apart in $\tilde{\mathbf{G}}^{k}_2$. 
The connections that were established to exist above are exactly the connections of the network constructed in Lemma \ref{lem:odd-node-connection}. By Lemma \ref{lem:odd-node-connection} it then follows that $\mathbf{G}_{k}$ satisfies $N_f$-robustness for all $k$ when $N_{k-1}$ is even.

Let $N_{k-1}$ be odd. Then $N_k$ is even. In this case, by construction the nodes that are $N_k/2$ links apart are connected. 
The links that were established to exist for the network $\mathbf{G}_k$ in the first paragraph and here is a sub-network of the construction given in Theorem \ref{thm:even-node-connection}, which was shown to be robust to removing any set of $N_f-1$ nodes in Theorem \ref{thm:alg-optimal-f} (please see the steps in the proof there). By Theorem \ref{thm:even-node-connection} it then follows that $\mathbf{G}_{k}$ satisfies $N_f$-robustness for all $k$ when $N_{k-1}$ is odd. 

Thus by induction, the $N_f$-robustness holds for all time steps because at time $k = 1$ we can construct the network according to Theorem \ref{thm:even-node-connection} or \ref{lem:odd-node-connection}, whichever is applicable. 

The networks by construction satisfy $\mathbf{G}_i \subseteq \mathbf{G}_j$ for all $i \leq j$. On arrival each new node is connected to $\tilde{N}_f$ other nodes within the same $\tilde{\mathbf{G}}^k_2$. Hence, at every step $\tilde{N}_f$ links are formed for each of the new nodes to connect it to the other nodes in the respective $\tilde{\mathbf{G}}^k_2$s. This corresponds to a total of $N_k\tilde{N}_f$ links by time step $k$. Further, when $N_k$ is even, links up to the limit of $N_k/2$ links are added (step (iv) of Theorem \ref{thm:sequence-odd-node-connection}). Hence,
\begin{align}
L_k & \leq N_k(N_f+1) + \sum_{l=1}^{N_k/2} l \nonumber \\
& \leq  N_k(N_f+1) + \frac{N_k}{4}\left(\frac{N_k}{2}-1\right) \nonumber \\
& \leq N_k(N_f+1) + \frac{N^2_k}{8}. \nonumber
\end{align}
When $N_k$ is odd, the summation is required only from $1$ to $(N_k-1)/2$. Hence,
\begin{align}
L_k & \leq N_k(N_f+1) + \sum_{l=1}^{(N_k-1)/4} l \nonumber \\
& \leq  N_k(N_f+1) + \frac{(N_k-1)}{4}\left(\frac{(N_k-1)}{2}-1\right) \nonumber \\
& \leq N_k(N_f+1) + \frac{(N_k-1)^2}{8}\ \blacksquare \nonumber
\end{align}

\section{Proof of Theorem \ref{thm:seq-dyn-alg-optimal-fl1}}
\label{sec:pf-Th10}

When exactly $n$ nodes are removed from one half and $N_k/2$ nodes from the other half, the only remaining nodes are the nodes from the first half. By construction each half is fully connected to start with and so the remaining nodes should be connected. Consider the case when less than $N_k/2$ nodes are removed from one half and more than $n$ from the other half. Clearly the remaining nodes in each half are connected because each half is fully connected to start with. Now, for every remaining node from the first half to not to be connected to any remaining node in the second half, the first $(n+1)$ nodes and the unique nodes they were connected to in the second half should have been removed. In that case, the number of nodes removed from the second half should be $n+1+\text{number of remaining nodes in the first half}$, which is strictly greater than $N_k/2+n$. Both cases together imply that, when $N_k/2+n$ nodes fail, the remaining network remains connected. The number of links is just the number of links in each of the halves plus the links crossing over from one half to the other. The number of links connecting nodes within each half count up to $N_k/4(N_k/2-1)$ and since there are two halves this adds up to $N_k/2(N_k/2-1)$. By construction the first $n+1$ nodes in each half are connected to every node in the other half. The remaining nodes in each half are connected to the first $n+1$ nodes in the other half and one more node in the other half. Hence, the total number of links, 
\begin{align} 
& L_k = N_k/2(N_k/2-1) + 2(N_k/2-(n+1))(n+2) \nonumber \\
& +(n+1)N_k \nonumber  \\
& = N_k/2(N_k/2-1) + 2N_k(n+3/2) - 2(n+1)(n+2) \nonumber \\
& = N^2_k/4 + 2N_k(n+3/4) - 2(n+1)(n+2)\ \blacksquare\nonumber 
\end{align}

\end{appendices}




\begin{IEEEbiography}[{\includegraphics[width=1in,height=1.25in,clip,keepaspectratio]{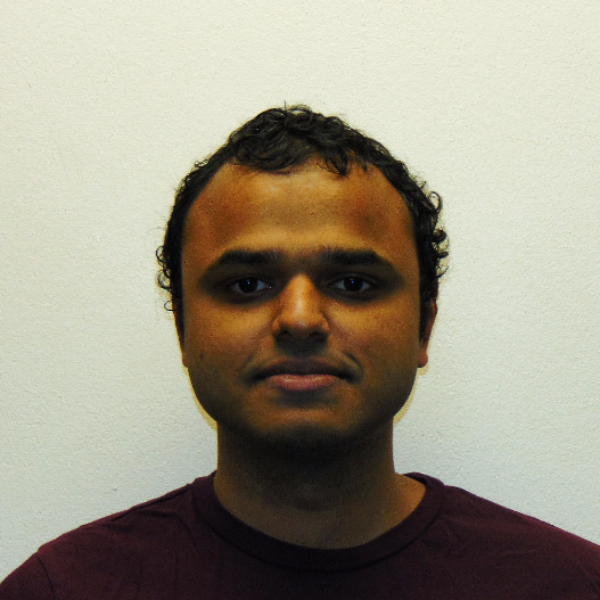}}]{Deepan Muthirayan}
is currently a Post-doctoral Researcher in the department of Electrical Engineering and Computer Science at University of California at Irvine. He obtained his Phd from the University of California at Berkeley (2016) and B.Tech/M.tech degree from the Indian Institute of Technology Madras (2010).  His current research interests include control theory, machine learning, topics at the intersection of learning and control, online learning, online algorithms, game theory, and their application to smart systems.
\end{IEEEbiography}

\vspace{-1em}

\begin{IEEEbiography}
[{\includegraphics[width=1in,height=1.25in,clip,keepaspectratio]{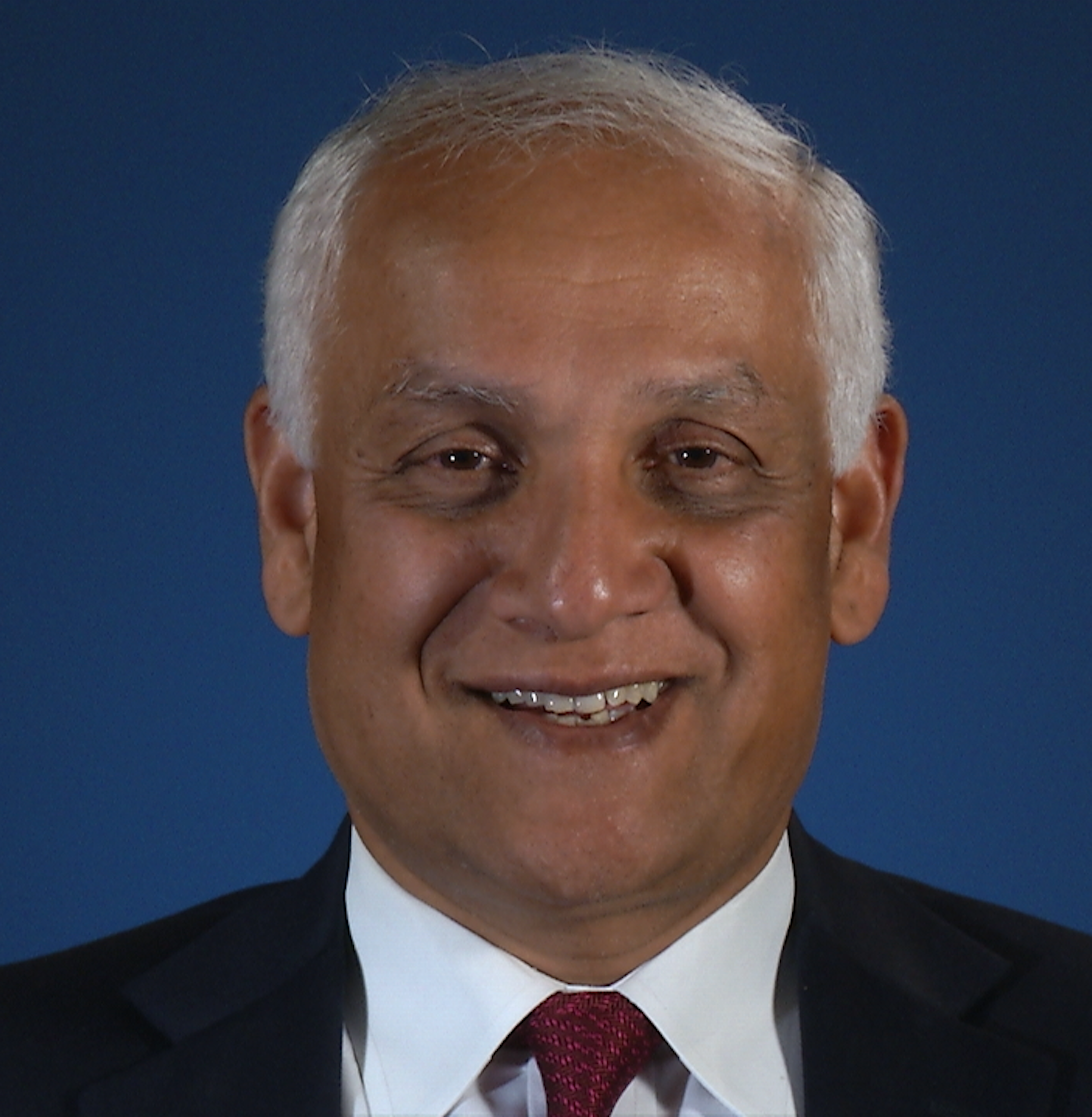}}]
{Pramod Khargonekar} received B. Tech. Degree in electrical engineering in 1977 from the Indian Institute of Technology, Bombay, India, and M.S. degree in mathematics in 1980 and Ph.D. degree in electrical engineering in 1981 from the University of Florida, respectively. Currently, he is Vice Chancellor for Research and Distinguished Professor of Electrical Engineering and Computer Science at the University of California, Irvine. His research and teaching interests are centered on theory and applications of systems and control. 
\end{IEEEbiography}

\end{document}